\documentclass[a4paper,11pt]{article}
\pdfoutput=1 % if your are submitting a pdflatex (i.e. if you have
\usepackage{jheppub} % for details on the use of the package, please
\usepackage[T1]{fontenc} % if needed
\usepackage{amsmath}
\usepackage{amsfonts}
\usepackage{amssymb}
\usepackage{graphicx}
\usepackage{hyperref}
\usepackage[normalem]{ulem}
\usepackage{multicol}
\usepackage{tcolorbox}
\usepackage{breqn}

\usepackage{physics}
\usepackage{xfrac}

\usepackage{array}
\usepackage{tabularx}

\newcommand{\mphi}{m_\phi}

\title{Probing displaced (dark)photons from low reheating freeze-in at the LHC}
\author[a]{Paola Arias,}
\author[b,c]{Bastián Díaz Sáez,}
\author[d]{Lucía Duarte,}
\author[e]{Joel Jones-Pérez,}
\author[e]{Walter Rodriguez,}
\author[e]{Danilo Zegarra~Herrera.}

\affiliation[a]{Facultad de Ingeniería, Universidad San Sebastián,
Santiago 8420524, Chile.}
\affiliation[b]{Millennium Institute for Subatomic Physics at the High-Energy Frontier (SAPHIR), Fernández
Concha 700, Santiago, Chile.}
\affiliation[c]{Instituto de Física, Pontificia Universidad Católica de Chile,
Avenida Vicuña Mackenna 4860, Santiago, Chile.}
\affiliation[d]{Instituto de Física, Facultad de Ciencias, Universidad de la República
Iguá 4225,(11400) Montevideo, Uruguay.}
\affiliation[e]{Sección Física, Departamento de Ciencias, Pontificia Universidad Católica del Perú, Apartado 1761, Lima, Perú.}

\emailAdd{paola.arias@uss.cl}
\emailAdd{bastian.diaz@uc.cl}
\emailAdd{lucia.duarte@fcien.edu.uy}
\emailAdd{jones.j@pucp.edu.pe}
\emailAdd{walter.rodriguez@pucp.edu.pe}
\emailAdd{danilo.zegarra@pucp.edu.pe}

\abstract{We extend the Standard Model (SM) by introducing a $U(1)'$ gauge boson and a real pseudo-scalar field, both odd under a $\mathbb{Z}_2$ symmetry. The resulting low-energy spectrum consists of a stable vector as the dark matter candidate, and a pseudo-scalar mediator, which interacts with the SM via a Higgs portal coupling and a dimension-five portal connecting it to both the dark and visible photons. We explore the freeze-in of both particles at low reheating temperature, finding a rich yield evolution dynamics in the early Universe. This setup brings a consistent dark matter scenario in which the dark photon relic abundance is generated through freeze-in at low reheating temperatures. In addition to its cosmological viability, the model can be tested at the LHC: Higgs bosons can decay into dark photons and displaced visible photons via the long-lived mediator. These signatures allow us to constrain the Higgs portal coupling using recent searches for non-pointing photons and limits on invisible or undetected Higgs decays. We derive meaningful constraints on the dark matter parameter space, in particular excluding a thermalized mediator in the region compatible with the observed relic abundance.}

\begin{document} 

\maketitle

\flushbottom

\section{Introduction}

The early history of the Universe and the nature of dark matter (DM) remain deeply mysterious. Observationally, Big Bang Nucleosynthesis (BBN) represents the earliest epoch that can be directly probed, thanks to precise measurements of the primordial abundances. These observations offer a robust window into the thermal history of the Universe at temperatures of a few MeV~\cite{Olive:1989xf,Copi:1994ev}. However, earlier stages remain largely unconstrained, allowing for the possibility of non-standard cosmological scenarios~\cite{Kawasaki:2000en,Hannestad:2004px,Hasegawa:2019jsa,Allahverdi:2020bys}. As a result, many ideas have been proposed that modify the standard picture between the end of inflation and BBN. In particular, DM production in such non-standard cosmologies can exhibit features that differ significantly from traditional expectations. Examples include an early matter-dominated era, and scenarios with a low reheating temperature $T_{\rm RH}$ --- defined as the temperature at which the post-inflationary Universe transitions into the standard hot Big Bang phase --- among many other possibilities~\cite{Giudice:2000ex,Fornengo:2002db,Pallis:2004yy,Gelmini:2006pw,Yaguna:2011ei,Drees:2017iod,Bernal:2018kcw,Arias:2019uol,Silva-Malpartida:2023yks,Silva-Malpartida:2024emu}.

On another front, dark matter (DM) models in which the dark sector is secluded from the Standard Model (SM) and interacts only via mediator particles --- through so-called portals --- have attracted considerable attention. The freeze-in mechanism~\cite{Hall:2009bx} is a commonly invoked framework to generate such secluded dark sectors, characterized by the fact that DM never reaches thermal equilibrium with the visible sector. Traditionally, this non-thermal production has been studied under the assumption that the maximum temperature involved, $T_{\rm RH}$, lies well above all relevant mass scales in both the SM and the dark sector. As a result, extremely small couplings between the two sectors are typically required, making their experimental detection highly challenging.

However, in recent years, the freeze-in production of DM at low $T_{\rm RH}$ has gained significant attention. This interest is driven by the realization that such cosmological scenarios are not only viable but also open new opportunities for laboratory tests. If $T_{\rm RH}$ lies below the masses of some dark-sector states, the corresponding coupling constants must increase to compensate for the Boltzmann suppression in the production rate. As a result, these models are testable not only in colliders but also in direct detection experiments~\cite{Roszkowski:2014lga,Bhattiprolu:2022sdd,Cosme:2023xpa, Koivunen:2024vhr, Arcadi:2024wwg, Boddy:2024vgt, Arcadi:2024obp,Barman:2024nhr,Barman:2024tjt,Lebedev:2024mbj, Bernal:2024ndy, Lee:2024wes, Belanger:2024yoj,Arias:2025nub, Borah:2025ema, Mondal:2025awq, Khan:2025keb, Bernal:2025qkj}. Remarkably, reheating temperatures as low as $\mathcal{O}(4~\mathrm{MeV})$ remain consistent with BBN~\cite{Kawasaki:2000en,Giudice:2000ex, Hannestad:2004px, Hasegawa:2019jsa,Allahverdi:2020bys}, making them cosmologically well motivated~\cite{Lebedev:2022cic}.

On the other hand, experimental collaborations at colliders have made remarkable efforts to improve their sensitivities to long-lived particles (LLPs), that is, particles with a lifetime that is long enough for them to travel a macroscopic distances before decaying~\cite{Curtin:2018mvb,Lee:2018pag,Alimena:2019zri}. This can lead to a series of non-standard signatures, such as displaced vertices, disappearing tracks, emerging jets and displaced photons. In particular, it is possible for the Higgs boson itself to decay into such states (see e.g.\ \cite{Strassler:2006im,Strassler:2006ri,Curtin:2013fra,Cui:2014twa,Craig:2015pha,Gago:2015vma,Caputo:2017pit,Jones-Perez:2019plk}).

Among the scenarios mentioned above, we highlight a case where the LLP can produce displaced visible photons in association with invisible particles, resulting in a non-pointing photon plus missing transverse energy signal at the LHC. This was first studied in the context of gauge mediated supersymmetry breaking models, with the lightest neutralino decaying into a photon and a gravitino~\cite{ATLAS:2013etx, ATLAS:2014kbb}. Then, it was found it could probe the decays of heavy neutral leptons and DM with dipole interactions~\cite{Primulando:2015lfa, Duarte:2016caz, Delgado:2022fea}. However, the sensitivity reach of the search for non-pointing photons coming from the decays of pair-produced LLPs from Higgs bosons, with a dedicated lepton trigger~\cite{ATLAS:2022vhr}, opened the possibility to constrain this kind of exotic Higgs decays, and place stringent bounds on the Higgs interaction with the LLPs~\cite{Duarte:2023tdw}. These studies have been complemented by several proposals to search for photons from LLP decays in future facilities~\cite{Barducci:2022gdv, Barducci:2024nvd,Bertuzzo:2024eds}, or with new proposed search strategies~\cite{Beltran:2024twr,Kim:2025tuz}. 

Interestingly, it has been shown that models featuring LLPs can also provide DM candidates, produced via freeze-in with low $T_{\rm RH}$, where the LLP acts as a mediator~\cite{Co:2015pka, Belanger:2018sti, Calibbi:2021fld, Bernal:2025qkj}. Specifically, it was shown that LHC searches for LLPs could place significant constraints on such models. Nevertheless,  the possibility of having an LLP mediator coupling to both photons and DM without reaching thermalization was not considered in detail in any of these previous studies. Thus, in this work we intend to bridge the gap, presenting a DM model with the relevant couplings, capable of generating the observed relic abundance, and demonstrating the capacity of displaced photons searches for applying bounds on the corresponding parameter space.

The paper is structured as follows. In the next section~\ref{sec:model} we introduce the details of our model and specify the relevant parameter space. Then, in section~\ref{sec:relic} we present the Boltzmann equations, and describe DM production in a simplified framework. We also introduce cosmological bounds from BBN and structure formation to constrain the DM parameter space. In section~\ref{sec:pheno} we review the search for displaced photons, as well as other relevant probes for LLPs, and describe how they can constrain the model. Finally, we conclude in section~\ref{sec:conc}.

\section{Model}\label{sec:model}

In this work, we consider a dark sector consisting of a new real pseudo-scalar field $\phi$ and a dark vector boson $A'_\mu$, the latter arising from a local $U'(1)$ symmetry. Both fields are assumed to be massive, with the mass of $A'_\mu$ generated either via a Stueckelberg mechanism or through a dark Higgs mechanism.\footnote{For an explicit realization of a dark Higgs mechanism generating a mass for a $\mathbb{Z}_2$-odd $A'_\mu$, see e.g.~\cite{Farzan:2012hh}.} We impose a $\mathbb{Z}_2$ symmetry under which the dark fields $A'_{\mu}$ and $\phi$ are odd, while all SM fields are even. This symmetry plays a crucial role in shaping the structure of the dark sector: it constrains the allowed couplings between the dark fields and prevents some direct interactions with the SM fields. In particular, it forbids kinetic mixing between the new $U'(1)$ gauge boson and the SM hypercharge gauge boson. As a result, the renormalizable part of the Lagrangian is extended by:
\begin{eqnarray}\label{lag1}
 \mathcal{L} \supset - \frac{1}{4}F'_{\mu\nu}F^{'\mu\nu} + \frac{1}{2}m_{\gamma'}^2A_\mu^{'2} + \frac{1}{2}(\partial_\mu \phi)^2 - \frac{1}{2}\tilde{m}_\phi^2\,\phi^2 - \lambda_\phi\,\phi^4 - \lambda_{HS}\,\phi^2 |H|^2,
\end{eqnarray}
where $H$ is the Higgs doublet of the SM. We take $(\lambda_\phi,\,\lambda_{HS})>0$ so that the scalar potential is bounded from below, and we ensure that $\mathbb Z_2$ remains unbroken. After electroweak symmetry breaking, $\phi$ acquires a new contribution to its mass, such that 
\begin{eqnarray}
  m_\phi^2 = \tilde{m}_\phi^2 + \frac{1}{2}\lambda_{HS}\,v_h^2.
\end{eqnarray}
with $v_h = 246$ GeV, the Higgs vacuum expectation value.

A key feature of this work will be the assumption of a decoupled sector --- in the EFT sense --- providing interaction terms for $A'_\mu$ via higher-dimensional operators. The only operator at $d=5$ involving $A'_\mu$ that does not break neither $U'(1)$ nor the $\mathbb Z_2$ symmetry is:
\begin{eqnarray}
\label{eq:portal}
\mathcal L_{5}= \frac{g_D}2 \phi\, F'_{\mu\nu}\tilde B^{\mu\nu},
\end{eqnarray}
where $g_D$ is a dimensionful parameter, related to a high-energy scale, $g_D = 1/\Lambda$, and $\tilde B^{\mu\nu}=\frac{1}{2}\epsilon^{\mu\nu\rho\lambda}B_{\rho\lambda}$ is the dual field strength of $U(1)_Y$. Considering that the model has four free parameters\footnote{From the collider perspective, $\lambda_\phi$ does not play a relevant role, so we do not define any specific target range. However, it may have an important impact on the cosmological evolution of the particle content; we therefore comment on it in the next section. } $(m_{\gamma'}, \, m_\phi, \,\lambda_{HS},\, g_D)$, in the following we identify the relevant parameter space based on the viability of having $\phi$ as a potential LLP.

\subsection{Target parameter space}\label{sec:parameterspace}
As mentioned in the introduction, we are interested in exploring the parameter space region of the model that can be accessed in collider experiments: more specifically, we want to reinterpret the results from displaced photon searches already carried out at the LHC ~\cite{CMS:2012bbi,ATLAS:2013etx,ATLAS:2014kbb,CMS:2019zxa,ATLAS:2022vhr} in terms of the relevant parameters of our model. 

Indeed, the most recent search in ATLAS~\cite{ATLAS:2022vhr} looks for neutral long-lived particles produced from Higgs boson exotic decays, which eventually disintegrate into photons and missing energy. In our context, if $m_\phi<m_h/2$, the Higgs can decay into $\phi$ pairs via the Higgs portal, that is, the last term in eq.~(\ref{lag1}) with $\lambda_{HS}$ coupling. Furthermore, if $m_{\gamma'}<m_\phi$, the pseudo-scalar becomes unstable, and decays into a photon and a dark photon via the dark photon portal, eq.~(\ref{eq:portal}), with $g_D$ coupling. This gives us the collider signal $p p \to h \to \phi \phi$, with each pseudo-scalar decaying as $\phi \to \gamma ~\gamma'$, which is exactly the signature probed by~\cite{ATLAS:2022vhr}. Therefore, in this work we will assume the hierarchy $m_{\gamma'}<m_\phi<m_h/2$, with the dark photon $\gamma'$ playing the role of stable DM.\footnote{Notice that the inverse hierarchy $m_\phi < m_{\gamma'}$ is a completely viable dark matter scenario (e.g. see \cite{Arias:2025nub}), with the $\phi$ being the DM candidate and $\gamma'$ being the unstable dark state.}
 
For displaced photon searches to be sensitive to our model, we need the scalar
to travel enough before decaying: 1~cm $\lesssim c\,\tau_\phi \lesssim 100$~m.  In addition, the displaced photon in the search must satisfy a cut on its transverse momentum, in fact, the search shows its maximum sensitivity when the particles producing the missing energy --- in our case, the dark photons --- are light enough to have the hardest possible visible photons. Based on previous work recasting this search~\cite{Duarte:2023tdw}, we focus on relatively light DM, with $m_{\gamma'}\lesssim 5$~GeV.

\begin{figure}[t!]
\centering
\includegraphics[width=0.6\textwidth]{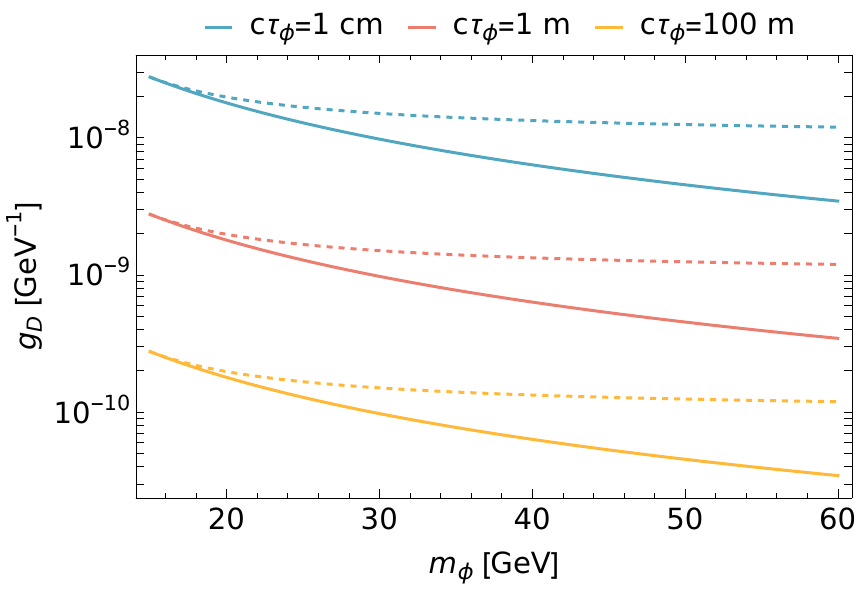}
\caption{Relevant parameter space to recast displaced photon searches. Cyan, red and orange curves correspond to decay lengths of 0.01, 1, and 100~m, respectively. Solid (dashed) lines show $m_{\gamma'}=0$ ($m_{\gamma'}=m_{\phi}-15$)~GeV. }
\label{fig:parameter_space}
\end{figure}
In order to understand which values of the dark photon portal coupling $g_D$ satisfy our requirement on $c\,\tau_\phi=c\,\hbar/\Gamma_\phi$, we consider the partial decay width:
\begin{eqnarray}\label{decay_width}
    \Gamma_\phi(\phi\to\gamma\gamma') = \frac{g_D^2\, c_W^2}{32\pi}m_\phi^3\left(1 - \frac{m_{\gamma'}^2}{m_\phi^2}\right)^3~,
\end{eqnarray}
where $c_W$ is the cosine of the weak mixing angle. Moreover, since the decay into a three body final state is suppressed by the off-shell $Z$ boson, we can safely take eq.~(\ref{decay_width}) as the dominant channel. This allows us to write the $\phi$ decay length as: 
\begin{eqnarray}
    c\,\tau_{\phi} \approx 
    (1.43~{\rm m})\left(\frac{3\times10^{-10}\,{\rm GeV}^{-1}}{g_D}\right)^2 \left(\frac{m_\phi}{60~{\rm GeV}}\right)^3\left(\frac{3500~{\rm GeV}^2}{\Delta m^2_{\phi\gamma'}}\right)^3 ,
    \label{eq:lifetime}
\end{eqnarray}
where $\Delta m^2_{\phi\gamma'}=m^2_\phi-m^2_{\gamma'}$ is compared to a benchmark value for $m_{\gamma'}=10$~GeV and $m_{\phi}=60$~GeV. In order to illustrate this result, we show in figure~\ref{fig:parameter_space} values of $c\,\tau_\phi$ in the $g_D-m_\phi$ plane. The curves correspond to $c\,\tau_\phi=0.01,\,1,\,100$~ m, with vanishing values of $m_{\gamma'}$ (solid lines), or equal to $m_{\phi}-15$~GeV (dashed lines). 
As discussed in section~\ref{sec:pheno}, the displaced photon search is most sensitive to LLP displacements around  $c \, \tau_\phi \approx 1$~m. 
Therefore, the figure suggests that the relevant parameter space lies within $g_D \sim 10^{-10}$ –- $10^{-9}$~GeV$^{-1}$.
Notice that, due to the $ Z \phi \gamma'$ interaction induced by the dark photon portal in eq.~(\ref{eq:portal}), the $g_D$ coupling is constrained by LEP measurements of the $Z$ boson invisible width and $Z \to \gamma +  {\rm inv.}$, depending on whether the pseudo-scalar decays outside the detector or not. In~\cite{Jodlowski:2024lab} the bound from LEP (in the $m_{\gamma'} \ll m_{\phi}\lesssim 60$~GeV limit) is found to require $g_D \lesssim 10^{-4}$~GeV$^{-1}$, far above our target scenario.

In contrast, the partial width for the decay $h\to\phi\phi$ is:
\begin{equation}
 \label{eq:GammaHiggs}
 \Gamma(h\to\phi\phi)=\frac{\lambda_{HS}^2\,v^2}{8\pi\, m_h}\sqrt{1-4\frac{m_\phi^2}{m_h^2}}~.
\end{equation}
Since a relatively large branching ratio is desirable to produce a sufficient number of scalars at the LHC, this typically calls for a sizable Higgs portal coupling, i.e., $\mathcal{O}(10^{-5}) \lesssim \lambda_{HS} \lesssim \mathcal{O}(1)$.
To simultaneously achieve a sizable $\phi$ pair-production cross section from Higgs decays, ensure that $\phi$ behaves as a long-lived particle (LLP) decaying within the inner detectors of the LHC, and maximize the sensitivity of non-pointing photon searches, we identify a compact region of parameter space that satisfies these key conditions:
\begin{itemize} \label{par1}
    \item $m_{\gamma'}\sim 1 ~{\rm{GeV}} \ll m_{\phi} \lesssim m_h/2$\,
    \item $10^{-11}$ GeV$^{-1}$ $\leq g_D \leq 10^{-9}$ GeV$^{-1}$\, 
    \item $10^{-5} \lesssim  \lambda_{HS} \lesssim 10^{-1}$
\end{itemize}
 Regarding the values of $\lambda_{HS}$, some regions are subject to existing constraints from Higgs data, which we will discuss in section \ref{sec:pheno}. Nonetheless, we include this broader range in section \ref{sec:relic}, for illustrative purposes.  Concerning the coupling $g_D$ our target parameter space remains unexplored in the mass range we propose to scan. On the other hand, stronger constraints apply in the sub-MeV range (e.g. \cite{Arias:2020tzl, Hook:2023smg}).

Having determined the parameter space to be studied based on collider phenomenology, we now turn to describe the specific DM production mechanisms for the dark photon and the pseudo-scalar particle.

\begin{figure}[t!]
\centering
\includegraphics[width=\textwidth]{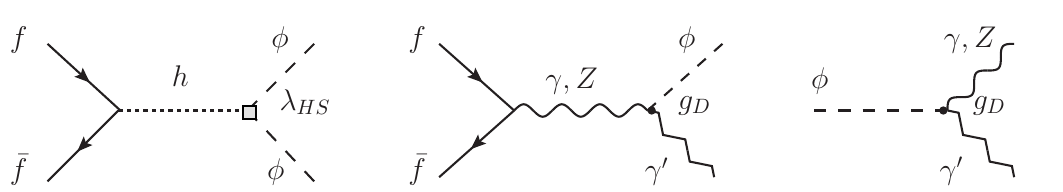}
\caption{Leading tree-level processes contributing to the production of the dark states, and the (inverse) decay relevant for freeze-in mechanism at low $T_{\rm RH}$.}
\label{fig:conversion-diagrams}
\end{figure}

\section{Dark matter production and cosmology}\label{sec:relic}

For our purposes, we assume the maximum temperature at which the dark sector particles are produced is the so-called reheating temperature $T_{\rm RH}$.\footnote{Reheating is not expected to be truly an instantaneous process, therefore, the maximum temperature can be higher than $T_{\rm RH}$, see e.g. \cite{Giudice:2000ex,Cosme:2024ndc}.} Even when other production processes may be present before reheating, we assume that before this period there is no population of $\gamma'$ nor $\phi$. We also assume an \textit{instantaneous} reheating epoch, as the SM sector is expected to thermalize rapidly due to its internal gauge couplings, after which the SM scattering and decays produce the dark sector from $T_{\rm RH}$ to lower temperatures.

Furthermore, we focus on scenarios with low reheating temperatures such that $T_{\rm RH}$ is comparable to or smaller than the new mass scales $m_{\gamma'}$ and $m_\phi$. In this regime, Boltzmann suppression efficiently reduces the production rates, enabling the correct relic abundance to be achieved via freeze-in. This consideration proves of high importance since, as shown in previous works, the 
UV-freeze-in mediated by $g_D$, with high reheating temperatures in the parameter space identified at the end of section~\ref{sec:parameterspace}, typically leads to DM overproduction (see e.g.\ figure~3 of~\cite{Arias:2025nub}). An alternative possibility is a fully thermal scenario \cite{DiazSaez:2024dzx}; however, it has been shown that this also leads to a significant overabundance in our target parameter space. In light of this, we explore the freeze-in production of dark states in the regime $m_{\gamma'} \lesssim T_{\rm RH} \ll m_\phi$. Specifically, as shown later, to fulfill the correct relic abundance in the parameter space of our interest, we must have $T_{\rm RH} \gtrsim \mathcal{O}(1\, \rm{GeV})$.

In this way, we establish a set of Boltzmann equations (BE) for the yields of $\gamma'$ and $\phi$ considering the freeze-in with low $T_{\rm RH}$ scales. For the parameter space identified in the last section, there are two relevant types of contribution to the freeze-in: 
 \begin{itemize}
  \item $f\bar{f}\rightarrow \phi\phi$,
  \item $f\bar{f}\rightarrow \phi\gamma'$,
 \end{itemize}
mediated by $\lambda_{HS}$ and $g_D$, respectively. Once dark states acquire non-negligible abundances, processes such as $\phi \leftrightarrow
  \gamma'\gamma$ may impact the yield of the population already produced of $\gamma'$ (see figure~\ref{fig:conversion-diagrams}).\footnote{We do not consider the processes $t\bar{t} \rightarrow \gamma'\phi,\, h \rightarrow \phi\phi, \,hh \rightarrow \phi\phi,\,W^+W^- \rightarrow \gamma'\phi,\, \phi h \rightarrow \gamma'\gamma,\, f\phi \rightarrow f\gamma'$, and $\gamma\gamma \rightarrow \gamma'\gamma' \,(\phi\phi)$, as their amplitudes are subleading in our target parameter space. The first five processes are strongly Boltzmann suppressed, and using {\tt micrOMEGAs}, we have verified that dark Primakov processes are also subdominant. The last process scales as $g_D^4$.} In addition, because the Higgs portal $\lambda_{HS}$ can be relatively large, there are scenarios where $\phi$ may thermalize, thereby introducing new features to the yield evolution.
  To avoid self-production and cannibalization processes, in the following analysis, we assume that $\lambda_\phi$ is small enough not to affect the relic abundance through dark freeze-out (see e.g., \cite{Bernal:2015xba, Heikinheimo:2017ofk,  Arcadi:2019oxh, Bernal:2020gzm} for related discussions).
 
In the following, we describe in detail the BE and its initial conditions, along with some solutions at low $T_{\rm RH}$.

\subsection{Boltzmann equations}
 To track the abundances of the dark states, we must consider all the relevant processes that may change the number of $\gamma'$
and $\phi$. Introducing the yields $Y = n/s$, with $n$ the number density of particles and $s = \frac{2\pi^2}{45}g_{*s}T^3$ the entropy density, the coupled BE for the dark states are given by\footnote{The effects of entropy injection and its consequences for the early Universe are considered in a following subsection, so we do not include a Boltzmann equation for photons. On the other hand, we have checked that the dilution of the DM abundance is negligible in our target parameter space.}
\begin{eqnarray} \label{beq1}
  \label{eq:beq1}\frac{dY_{\gamma'}}{dT}  &=& -\frac{1}{\overline{H} T s}\left(\sum_f C_{f\bar{f}\leftrightarrow \phi\gamma'} + C_{\phi\leftrightarrow\gamma\gamma'}\right), \\ \label{beq2}
      \frac{dY_{\phi}}{dT}  &=& -\frac{1}{\overline{H} T s}\left(\sum_f C_{f\bar{f}\leftrightarrow \phi\gamma'} + \sum_f C_{f\bar{f}\leftrightarrow \phi\phi}  - C_{\phi\leftrightarrow\gamma\gamma'}\right),  
 \end{eqnarray}
 where $H/\overline{H} \equiv 1 + \frac{T}{3}\frac{d\log g_{*s}}{dT}$,  the Hubble rate is given by $H = \sqrt{\frac{\pi^2g_*}{90}}\frac{T^2}{M_P}$, where $g_*(T)$ is the effective number of relativistic degrees of freedom at temperature $T$, and $M_P = 2.4\times 10^{18}$~GeV is the reduced Planck mass. The $C$ factors in the right side of each BE correspond to integrated collision terms, with $C_{A\leftrightarrow B} \equiv C_{A\rightarrow B} - C_{B\rightarrow A}$, with
 \cite{Belanger:2018ccd} 
 \begin{multline}
 \label{coll_terms}
    C_{A\rightarrow B} = \int\prod_i \left(\frac{d^3p_i}{(2\pi)^3 2E_i} f_i\right)\prod_j \left(\frac{d^3p_j}{(2\pi)^3 2E_j} (1 \pm f_j)\right) \\
      \times (2\pi)^4\delta^4(p_1 + p_2 - p_3 - p_4)C_A \ev{|\mathcal{M}|^2},
 \end{multline} 
 where $f_i$ is the distribution phase-space function of the respective particle, and $C_A$ a combinatory factor 1/2 for identical initial states and 1 otherwise. 
 
 In the following, we make use of the Maxwell-Boltzmann equilibrium number density given by
 \begin{equation}\label{nie}
  n_{X,e} =  g_{X}\frac{m_{X}^2}{2\pi^2} ~T ~K_2\left(\frac{m_{X}}{T}\right),
 \end{equation}
 where $X$ represents a specific particle, $g_{X}$ its internal degrees of freedom, and $K_2$ the modified Bessel function of the second kind.

 Let us state more explicitly each collision term shown in the set of BEs presented in eqs.~(\ref{beq1}) and~(\ref{beq2}). 
 First, the collision term for the reaction $f\bar{f}\rightarrow \phi\gamma'$ in eq.~(\ref{coll_terms}) can be reduced to
 \begin{eqnarray}
 C_{f\bar{f}\rightarrow \phi\gamma'} = n_{f,e}^2 \ev{\sigma v}_{f\bar{f}\rightarrow  \phi\gamma'}, 
 \end{eqnarray}
 where we assume an equilibrium number density for the SM initial state fermions. 
The corresponding backreaction is given by
\begin{eqnarray}
    C_{\phi\gamma'\rightarrow f\bar{f}} &=& n_{\phi}n_{\gamma'} \ev{\sigma v}_{\phi\gamma'\rightarrow  f\bar{f}} .
\end{eqnarray}

Making use of the detailed balance principle, we express the collision terms in the following way,
\begin{eqnarray}
 C_{f\bar{f}\leftrightarrow \phi\gamma'} &=& n_{f,e}^2 \ev{\sigma v}_{f\bar{f}\rightarrow  \phi\gamma'} -  n_{\phi}n_{\gamma'} \ev{\sigma v}_{\phi\gamma'\rightarrow  f\bar{f}}, \nonumber \\
    &=& n_{\phi,e}n_{\gamma',e} \ev{\sigma v}_{\phi\gamma'\rightarrow  f\bar{f}} - n_{\phi}n_{\gamma'} \ev{\sigma v}_{\phi\gamma'\rightarrow  f\bar{f}}.  \label{C1}
\end{eqnarray}
Now, as we focus on the parameter space where $m_{\gamma'} \lesssim  T_{\rm RH} \ll m_\phi$, the annihilation $\gamma'\phi \rightarrow f\bar{f}$ occurs approximately in a non-relativistic way, i.e. at low relative velocities $v$. In this case, we can obtain $\ev{\sigma v}_{\phi\gamma'\rightarrow  f\bar{f}}$ by expanding $(\sigma v)_{\phi\gamma'\rightarrow  f\bar{f}}$
in powers of relative velocity of the incoming initial dark states, retaining the leading terms. Explicitly, 
\begin{eqnarray}\label{eq:expandsigv}
    \ev{\sigma v}_{\phi\gamma'\rightarrow  f\bar{f}} \approx (\sigma v)_{\phi\gamma'\rightarrow  f\bar{f}}|_{s\rightarrow (m_{\gamma'} + m_\phi)^2\left(1 + \frac{v^2}{4}\right)},
\end{eqnarray}
where $v^2 \rightarrow 6T/\mu$, with $\mu = m_{\gamma'}m_\phi/(m_{\gamma'}+m_\phi)$. The explicit expressions for each type of SM fermion in $\ev{\sigma v}_{\phi\gamma'\rightarrow  f\bar{f}}$ are given in appendix~\ref{app_a}.

On the other hand, the integrated collision terms from the Higgs portal are given by 
  \begin{eqnarray} \label{C2}
     C_{f\bar{f} \rightarrow \phi\phi} &=& n_{f,e}^2 \ev{\sigma v}_{f\bar{f} \rightarrow \phi\phi} = n_{\phi, e}^2 \ev{\sigma v}_{\phi\phi\rightarrow f\bar{f}}, \\
     \label{C2_back}C_{\phi\phi\rightarrow f\bar{f}} &=& n_{\phi}^2 \ev{\sigma v}_{\phi\phi\rightarrow f\bar{f}},
  \end{eqnarray}
  where in eq.~(\ref{C2}) we have used again detailed balance in order to express both collision terms as a function of $\ev{\sigma v}_{\phi\phi\rightarrow f\bar{f}}$. We are thus allowed to use the algebraic non-relativistic expansion of $\ev{\sigma v}_{\phi\phi\rightarrow ff}$ given by
  \cite{Cline:2013gha}
 \begin{eqnarray}\label{ff}
   \ev{\sigma v}_{\phi\phi\rightarrow f\bar{f}} = \frac{m_f^2\, \lambda_{HS}^2}{4\pi \left[(m_h^2 - 4m_\phi^2)^2 + m_h^2\, \Gamma_h^2\right]}X_f\left(1 - \frac{m_f^2}{m_\phi^2}\right)^{3/2},
 \end{eqnarray}
 where $\Gamma_h \approx 4$ MeV, and $X_f$ is a factor equal to 1 for leptons and $\approx 3$ for quarks, including one-loop QCD corrections (see appendix~A of \cite{Cline:2013gha}). Similarly to the case in eq.~(\ref{eq:expandsigv}) involving the dark photon portal for freeze-in, expression~(\ref{ff}) is useful for computational calculations.

 Finally, the decay and inverse decay collision terms for $\phi \leftrightarrow \gamma\gamma'$ are given by
 \begin{eqnarray}
     C_{\phi \leftrightarrow \gamma\gamma'} &=& C_{\phi \rightarrow \gamma\gamma'} - C_{\gamma\gamma'\rightarrow \phi} \nonumber \\
     &=& \Gamma_\phi\left(n_\phi - n_{\gamma'}\frac{n_{\phi,e}}{n_{\gamma',e}}\right), \label{eq_decay}
 \end{eqnarray}
 with $\Gamma_\phi$ given in eq.~(\ref{decay_width}).

 We have implemented the set of BEs considering the approximations presented above, and cross-checked the relevant cross sections and decay widths with {\tt micrOMEGAs}.
 Before presenting the numerical results, we emphasize the following. 
 Firstly, the results in the subsequent sections are only reliable within the parameter space we focus on; for instance, for much higher $T_{\rm RH}$, additional collision terms in the coupled Boltzmann equations—such as $W^+W^-$ or $hh$ annihilation—should be included (see e.g.\ \cite{Arias:2025nub}). Secondly, the non-relativistic expansion used to compute \( \ev{\sigma v}_{\gamma'\phi \rightarrow f\bar{f}} \) becomes invalid for $T_{\rm RH}\gg m_{\gamma'}$ as $\gamma'$ behaves relativistically in the cosmic frame.
 Notice that in the region of parameter space where the dark photon portal significantly contributes to the collision terms of the coupled Boltzmann equations (e.g., for relatively small values of $\lambda_{HS}$), the non-relativistic approximation can differ by up to a factor of two in the calculation of the relic abundance, compared to the standard result obtained by thermally averaging $\sigma v$~\cite{Gondolo:1990dk}.

 \subsection{Results}
 The two portals $g_D$ and $\lambda_{HS}$ leave us with a rich phenomenology of DM production in the early Universe. As stated above, we take the initial abundances of both dark particles to be negligible at $T_{\rm RH}$, while the SM is thermal. Due to the dark sector and SM interactions, both dark states are produced, but the yield evolution depends strongly on the model parameters and $T_{\rm RH}$. We now examine possible production regimes based on these dynamics.  
 %%%%%%%%%%%%%%%%%%%%%
  \begin{figure}[t!]
  \begin{center}   
\includegraphics[width=1\textwidth]{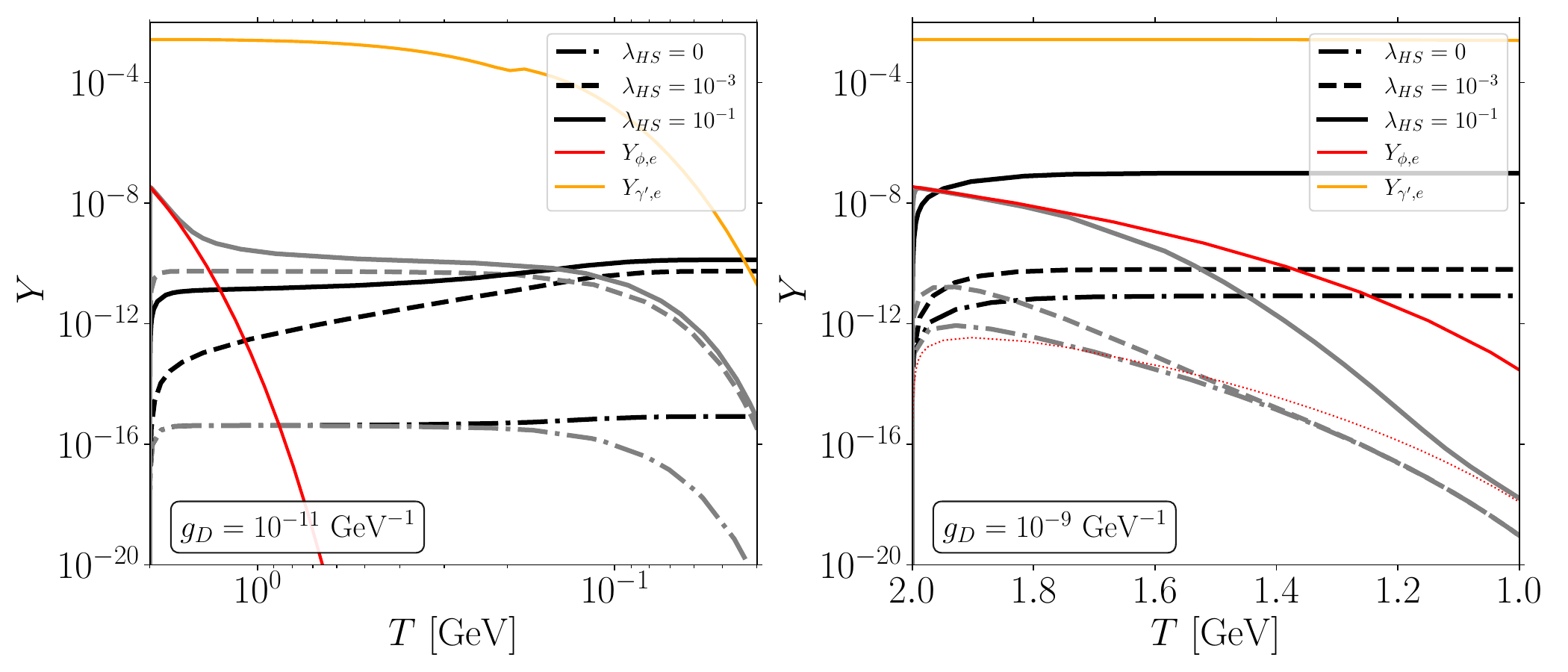}
\caption{Yield evolution of $\phi$ (gray lines) and $\gamma'$ (black lines) starting from $T_{\rm RH} = 2~\mathrm{GeV}$. In both plots, we fix $m_{\gamma'} = 1~\mathrm{GeV}$ and $m_\phi = 30~\mathrm{GeV}$. The red and orange solid lines correspond to the equilibrium yields, while the red dotted line in the right panel represents the combination $Y_{\gamma'} \, Y_{\phi,\mathrm{eq}} / Y_{\gamma',\mathrm{eq}}$ for the case $\lambda_{HS} = 0.1$. The red dotted line illustrates the quasi-static equilibrium~\cite{Hall:2009bx}, where $\phi$ participates via the process $\phi \leftrightarrow \gamma'\gamma$ before full chemical decoupling.
} \label{fig:3}  
  \end{center}  
      \end{figure}
%%%%%%%%%%%%%%%%%%%%%

We illustrate the freeze-in production of dark matter by solving the coupled Boltzmann equations~(\ref{beq1}) and~(\ref{beq2}), with vanishing initial yields for the dark sector particles. Figure~\ref{fig:3} shows the resulting yields of $\gamma'$ (black lines) and $\phi$ (gray lines) as functions of temperature, for the benchmark values: $T_{\rm RH} = 2~\mathrm{GeV}$, $m_{\gamma'} = 1~\mathrm{GeV}$, $m_\phi = 30~\mathrm{GeV}$, and two values of $g_D = 10^{-11}$ (left panel) and $g_D =10^{-9}~\mathrm{GeV}^{-1}$ (right panel).

We identify different behaviors depending on whether the mediator $\phi$ reaches thermal equilibrium or not. In all cases, $\gamma'$ never thermalizes with the plasma and is produced via freeze-in:

\begin{itemize}
    \item \emph{Non-thermal $\phi$:} For small values of $g_D$ (left panel), $\phi$ is long-lived and can accumulate a sizable abundance after reheating. Then it slowly converts into $\gamma'$ through $\phi \leftrightarrow \gamma'\gamma$ processes. This enhances the yield of $\gamma'$ until the crossover point where $n_\phi \approx (n_{\gamma'} \cdot n_{\phi,e})/n_{\gamma',e}$. For higher values of $g_D$ (right panel) the decay is faster and contributes only slightly to the DM relic abundance.

    \item \emph{Thermalized $\phi$:} For larger portal coupling $\lambda_{HS}$ (solid gray lines), $\phi$ can thermalize with the SM plasma ($Y_\phi \approx Y_{\phi,e}$). After thermal decoupling, it decays contributing to the relic abundance of $\gamma'$ (in this case, $\gamma'$ behaves as a super-WIMP~\cite{Feng:2003xh,Feng:2003uy,Asaka:2005cn,Asaka:2006fs,Molinaro:2014lfa,Faber:2019mti}). For small $g_D$ (left panel), this delayed production significantly enhances $Y_{\gamma'}$. For larger $g_D$ (right panel), the decay of $\phi$ is too fast to impact $\gamma'$ significantly.\footnote{In this regime, $\phi$ decouples from equilibrium before fully depleting, entering a so-called quasi-static equilibrium. This behavior is typical of mediators that decay faster than they annihilate, and may eventually freeze-out.}
\end{itemize}

\begin{figure}[t!]
  \begin{flushright}
    \makebox[\textwidth][l]{%
      \includegraphics[width=1.15\textwidth]{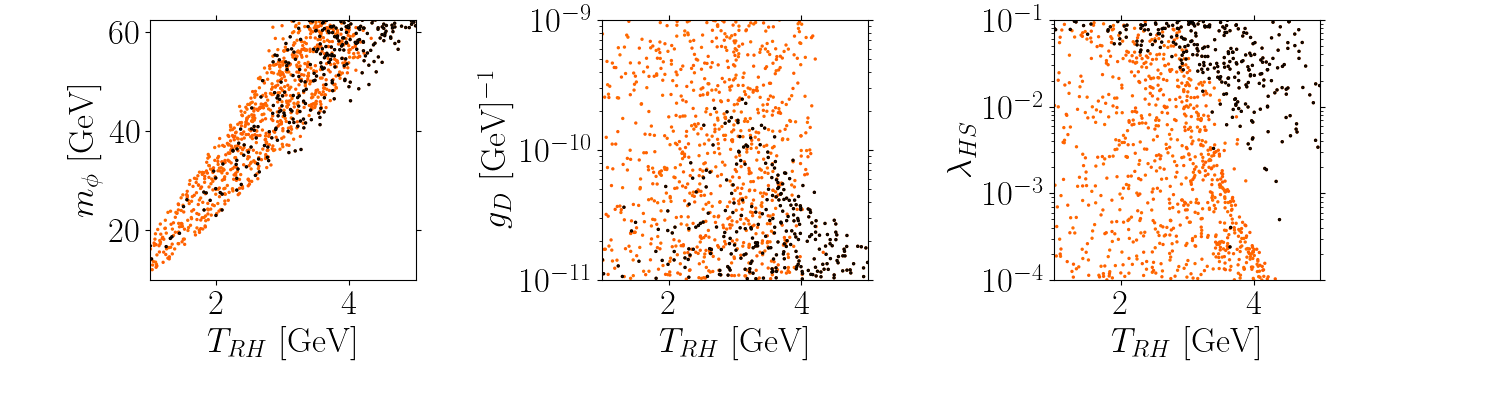}
    }
    \caption{Random scan points that yield the correct relic abundance for $m_{\gamma'} = 1$ GeV. Orange and black points indicate the regions where $\phi$ does not thermalize and where it does, respectively. Further details on the scan are provided in the text.}
    \label{plot:ran}
  \end{flushright}
\end{figure}

To illustrate the differences between the thermal and non-thermal $\phi$ scenarios in a more general way, in figure~\ref{plot:ran} we show the result of a random scan over a selected region of the parameter space for $m_{\gamma'} = 1$~GeV, retaining only the points that yield the correct relic abundance $\Omega_{\gamma'} h^2 = 0.12 \pm 0.02$. The scan is performed using a linear scale for $m_\phi \in [10, \, 62.5]$~GeV and $T_{\rm RH} \in [1,\, 5]$~GeV, and a logarithmic scale for $g_D \in [10^{-11}, \, 10^{-9}]$~GeV$^{-1}$ and $\lambda_{HS} \in [10^{-3}, \, 10^{-1}]$. In the figure, the black points correspond to the scenarios in which $\phi$ thermalizes and undergoes freeze-out, while for the orange points $\phi$ never enters in equilibrium with the SM.

From the first panel of figure~\ref{plot:ran}, a clear proportionality between $T_{\rm RH}$ and $m_\phi$ emerges. As $T_{\rm RH}$ decreases, the collision terms in eqs.~(\ref{C1}) and~(\ref{C2}) become increasingly Boltzmann suppressed, since they scale as powers of $e^{-m_\phi/T}$. To compensate for this suppression and still achieve the observed relic abundance, $m_\phi$ must also decrease.  This trend is reflected in the scarcity of thermalization points at low $T_{\rm RH}$, as the production rates become too suppressed for the mediator to equilibrate with the plasma. Secondly, note that large coupling constants do not necessarily imply thermalization, as shown by the plots in the middle and to the right of figure~\ref{plot:ran}. In general, a high $\lambda_{HS}$ is required, although also depending on other parameters, e.g. $g_D$ or $T_{\rm RH}$.

\begin{figure}[t!]
  \begin{center} 
  \includegraphics[width=0.45\textwidth]{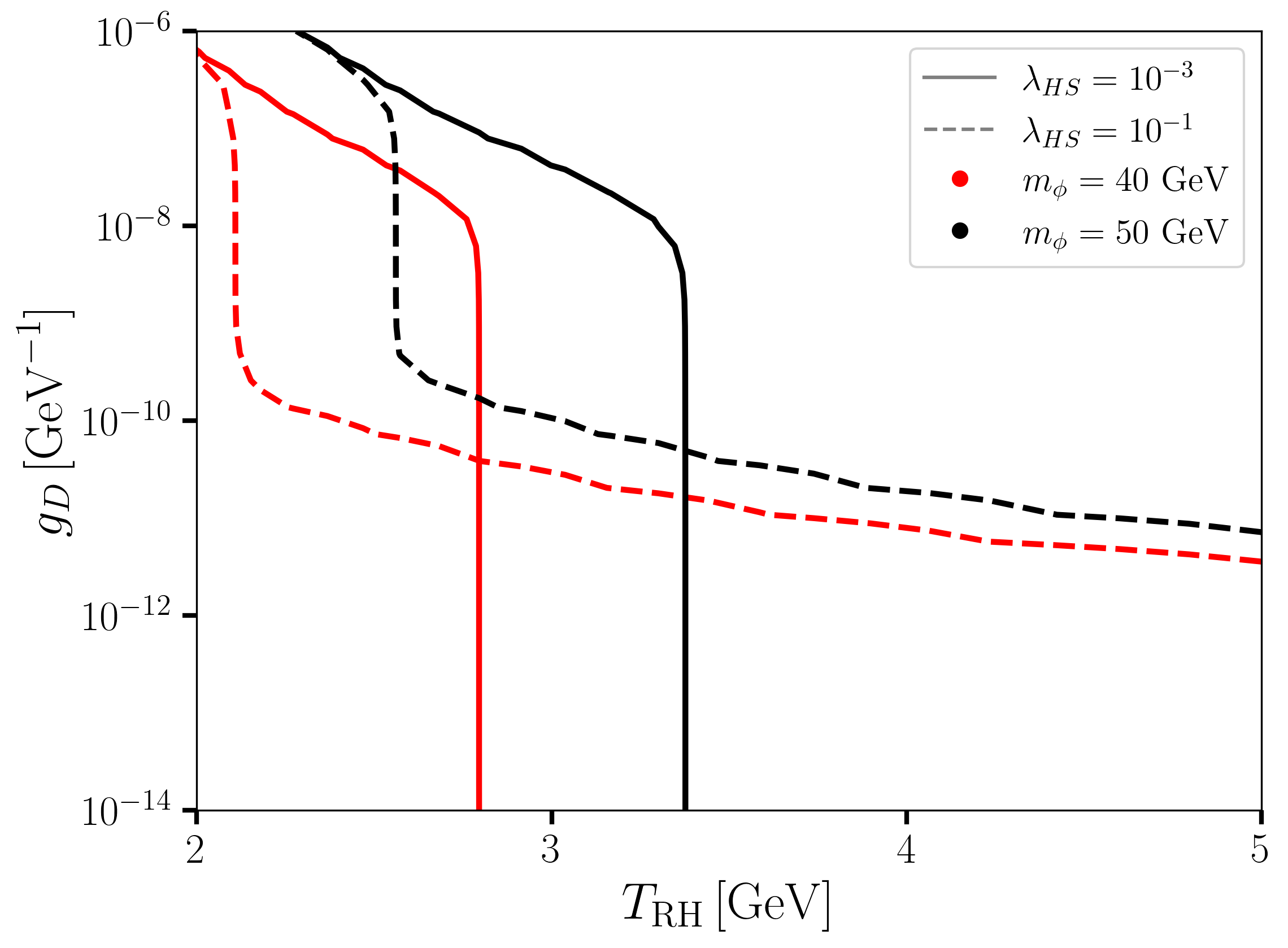}
\includegraphics[width=0.45\textwidth]{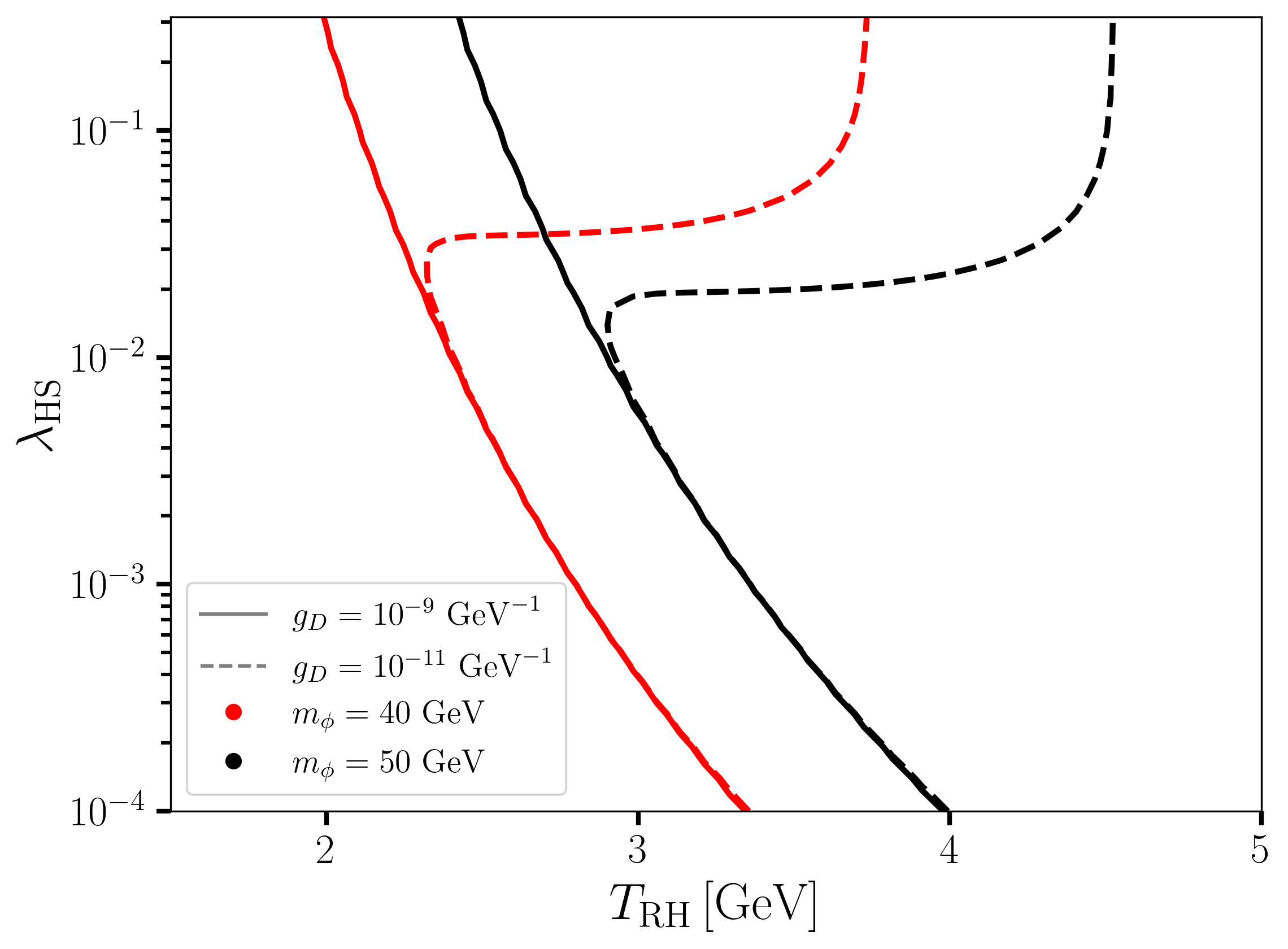}
   \caption{Portal couplings as a function of the reheating temperature for $\mphi = 40$~GeV (red) and 50~GeV (black), with $m_{\gamma'} = 1$~GeV. All curves reproduce the correct dark photon relic abundance. (Left) Solid and dashed lines use $\lambda_{ HS} = 10^{-3}$ and $10^{-1}$, respectively. (Right) Solid and dashed lines use $g_D = 10^{-9}$~GeV$^{-1}$ and $10^{-11}$~GeV$^{-1}$, respectively. }
   \label{plot:couplings_vs_TRH}
  \end{center}  
      \end{figure}

To complement the analysis, in figure~\ref{plot:couplings_vs_TRH} we show the contours of the correct relic abundance in the planes $T_{\rm RH}-\, g_D$ (left) and $T_{\rm RH}\,- \lambda_{HS}$ (right). For the left panel, the contours have $\lambda_{ HS}=10^{-3}$ (solid) and $10^{-1}$ (dashed). For the right one, the contours have $g_D=10^{-9}$~GeV$^{-1}$ (solid) and $10^{-11}$~GeV$^{-1}$ (dashed). Both panels have the dark photon mass fixed to $m_{\gamma'}=1$~GeV. We can recognize the different production mechanisms and their dependence on the interaction strength and the reheating temperature. Concerning the left panel of the figure:
\begin{itemize}
    \item For high $g_D$ and low $T_{\rm RH}$ (upper left corner), production is dominated by the freeze-in at strong regime, as the $\phi$ abundance is strongly suppressed. In this case, the Higgs portal plays no role in the dark matter production. As expected, the  behavior extends for small $\lambda_{HS}$ (solid lines).
    \item As $g_D$ decreases and $T_{\rm RH}$ increases, the freeze-in production of $\phi$ via the Higgs portal interaction becomes relevant, and dominates the production of $\gamma'$ indirectly. The vertical line in the $g_D$–$T_{\rm RH}$ plane is straightforward to interpret: since the dark masses and Higgs portal coupling are fixed, requiring the correct relic abundance determines $T_{\rm RH}$, independently of $g_D$ (which does not affect the yield in this regime). 
    \item For large $\lambda_{HS}$ (dashed lines), thermalization of the mediator is not guaranteed, as $\phi$ decay can be efficient even for moderately large $g_D$, disrupting its thermalization. When $g_D \lesssim 10^{-10}$~GeV$^{-1}$, thermalization becomes possible for $\lambda_{HS} = 0.1$ and thus $\phi$ freezes-out. This corresponds to the lower part of the curve for both $\mphi$ masses. A residual dependence on $g_D$ and $T_{\rm RH}$ remains, as higher $T_{\rm RH}$ enhances the decay contribution to the dark photon yield. In this regime the dark photon is produced through the so-called super-WIMP mechanism.
\end{itemize}
%%%%%%%%%%%%%%%%%%%%%%%%
%%%%%%%%%%%%%%%%%%%%
In the $\lambda_{HS}$–$T_{\rm RH}$ plane (right panel of figure~\ref{plot:couplings_vs_TRH}), we see that when $g_D$ is fixed, reproducing the correct relic abundance at larger $\lambda_{HS}$ requires lowering $T_{\rm RH}$. The super-WIMP production appears in the upper part of the plot ($\lambda_{HS} \sim 0.1$) as a plateau in $T_{\rm RH}$ for the dashed lines (corresponding to smaller $g_D$). In this regime, the $\phi$ yield is set by freeze-out, thus independent of $T_{\rm RH}$. However, as $T_{\rm RH}$ increases, $\phi$ decays efficiently while in equilibrium, interrupting the freeze-out and altering the relic abundance.

\subsection{Cosmological constraints of the parameter space}

The cosmological history of the present model is strongly influenced by the lifetime of $\phi$ and its abundance prior to decay. Particularly dangerous are decays occurring around the epoch of light element formation, BBN. On the one hand, electromagnetic decays can modify the baryon-to-photon ratio and disintegrate light nuclei. Additionally, if the decay occurs after neutrino decoupling, it can lower the neutrino temperature. On the other hand, decays into relativistic dark photons can impact BBN through contributions to the effective number of neutrino species. Moreover, we must ensure that the dark photon cools sufficiently before matter-radiation equality, so as not to interfere with the formation of large-scale structures in the early Universe. Since these constraints depend only on the $g_D$ coupling, they have already been studied in~\cite{Arias:2025nub}; nevertheless, we briefly review them here for completeness.      

\subsubsection{Dark matter warmness}
In order not to interfere with the formation of large structures in the Universe, the DM velocity is restricted to satisfy \cite{Viel:2005qj, Viel:2007mv,Baur:2015jsy,Chatrchyan:2020pzh} $v_{\rm DM}\lesssim 10^{-3} c, \, \mbox{at}\,\, a=a_{\rm EQ}$, 
with $a_{\rm EQ} $ the scale factor at matter-radiation equality. 
Dark photons from the decay of $\phi$ gain substantial momentum, given by
\begin{align}
k_{\gamma' \rm dec}=\frac{m_\phi^2-m_{\gamma'}^2}{2\mphi}.
\end{align}
Therefore, the physical momentum  redshifts according to $p_{\gamma'}=k_{\gamma' \rm dec}\,{a_{\rm dec}}/{a}$. 
The dark photon becomes nonrelativistic at $p_{\gamma'}\sim m_{\gamma'}$, from where we find $a_{\rm nr}=k_{\gamma' \rm dec}\,a_{\rm dec}/m_{\gamma'}$. Then, replacing in the DM's velocity expression $v=a_{\rm nr}/a$ we get
\begin{align}
v_{\gamma'}=\frac{k_{\gamma' \rm dec}}{m_{\gamma'}}\frac{a_{\rm dec}}a,
\end{align}
where imposing the velocity restriction and using $T\propto a^{-1}$, we find
\begin{align}
T_{\rm dec}> \frac{k_{\gamma' \rm dec}}{m_{\gamma'}}\, T_{\rm EQ} \,10^3.
\end{align}
Here, $T_{\rm EQ}\sim 0.8$~eV is the temperature at matter-radiation equality. 
This constraint can be more conveniently expressed in terms of the lifetime $\tau_{\phi}$ of the scalar, by considering $T_{\rm dec}/T_{\rm EQ}\sim \left(t_{\rm EQ}/\tau_\phi\right)^{1/2}$, obtaining
\begin{align}
\tau_{\phi}<1.9\times 10^5 \, \mbox{sec}\,\left(\frac{m_{\gamma'}}{10\,\mbox{GeV}}\right)^2 \left(\frac{\mphi}{60\,\mbox{GeV}}\right)^2 \left(\frac{3500\,\mbox{GeV}^2}{\Delta m_{\phi\gamma'}^2}\right)^2, 
\end{align}
where we considered $t_{\rm EQ}\sim 51 $~kyr. The lifetime of the scalar mediator has been already obtained in eq.~(\ref{eq:lifetime}) to be
\begin{eqnarray}
    \tau_{\phi} \approx 
    4.9\times 10^{-9}\,\mbox{sec}\,\left(\frac{3\times10^{-10}\,{\rm GeV}^{-1}}{g_D}\right)^2 \left(\frac{m_\phi}{60~{\rm GeV}}\right)^3\left(\frac{3500~{\rm GeV}^2}{\Delta m^2_{\phi\gamma'}}\right)^3
\end{eqnarray}
in our target parameter space ballpark. Therefore, we do not expect conflicts with structure formation.

\subsubsection{BBN constraints}
We now examine the conditions under which the cosmological evolution of our model could affect the BBN epoch and the effective number of neutrinos. The decay of the scalar mediator can influence cosmological observables depending on the temperature at which it occurs. The consequences of such decays can be categorized according to three temperature regimes: 

\begin{itemize}
\item  If the decay happens before BBN ($T_{\rm dec} \gtrsim 1$~MeV), the resulting dark photons are still relativistic and contribute to the radiation energy density. This alters the effective number of neutrino species $N_{\rm eff}$ during BBN, potentially increasing its value.

\item If, in addition, the decay occurs after neutrinos have decoupled, entropy is injected only into the photon bath. This reduces the neutrino-to-photon temperature ratio and leads to a smaller $N_{\rm eff}$ at the time of CMB decoupling.

\item If the decay takes place during the BBN era, it affects both the baryon-to-photon ratio and $N_{\rm eff}$ through entropy release and the presence of additional relativistic degrees of freedom.
\end{itemize}

A detailed treatment of the above scenarios would require solving the full set of BBN equations, including the evolution of light nuclei. Since such an analysis goes beyond the scope of this work, we instead rely on the results of Ref.~\cite{Yeh:2024ors}, where the authors study the cosmological impact of a general non-relativistic species $X$ decaying near the BBN epoch, accounting for both visible and invisible decay modes.

To quantify the effect, they introduce a parameter $\xi$, defined as the ratio between the energy density of the decaying species and that of radiation, evaluated at an early reference temperature of 10 MeV:
\begin{align}
\frac{\rho_X(T)}{\rho_r(T)}\Bigg|_{T = 10\,\text{MeV}} = \xi.
\label{eq:ratio_olive_rephrased}
\end{align}

The analysis combines likelihoods based on theoretical predictions for BBN (including observational inputs) and Planck CMB data, allowing one to constrain the model parameters: the lifetime $\tau_X = \Gamma_X^{-1}$, the baryon-to-photon ratio $\eta$, and the initial energy fraction $\xi$. By marginalizing over these parameters, bounds can be obtained for any desired subset.

In scenarios where the decay products are electromagnetic, Ref.~\cite{Yeh:2024ors} finds a mild preference for a non-zero energy injection, driven in part by a slight preference in CMB observations for $N_\nu < 3$. The preferred region corresponds to $\xi (\tau_X/1\,\text{sec})^{1/2} \sim 0.015$ for lifetimes between $1$ and $100$ seconds. On the other hand, in models with invisible (dark sector) decays, the strongest constraints arise from $N_{\rm eff}$, with BBN data setting the leading limits for short lifetimes and CMB data dominating at larger lifetimes.

Our scenario interpolates between these limiting cases. Decays into photons can lower $N_{\rm eff}$ and affect the baryon asymmetry, while invisible decays can enhance the radiation density. To ensure consistency with observational constraints, we conservatively adopt the bound $\xi \lesssim 10^{-5}$, assume an initial temperature of $T_{\rm ini} = 10$~MeV before BBN, and restrict the lifetime of $\phi$ to below $10^4$ seconds, which is totally safe for our purposes. This ensures that the decays occur before the Universe cools below $T = 10$~keV, thereby avoiding severe constraints from the photodisintegration of light elements \cite{Cyburt:2002uv, Hufnagel:2017dgo, Hufnagel:2018bjp, Kawasaki:2020qxm, Depta:2020zbh}. 
%%%%%%%%%%%%%%%%
\begin{figure}[t!]
\centering
\includegraphics[width=0.8\textwidth]{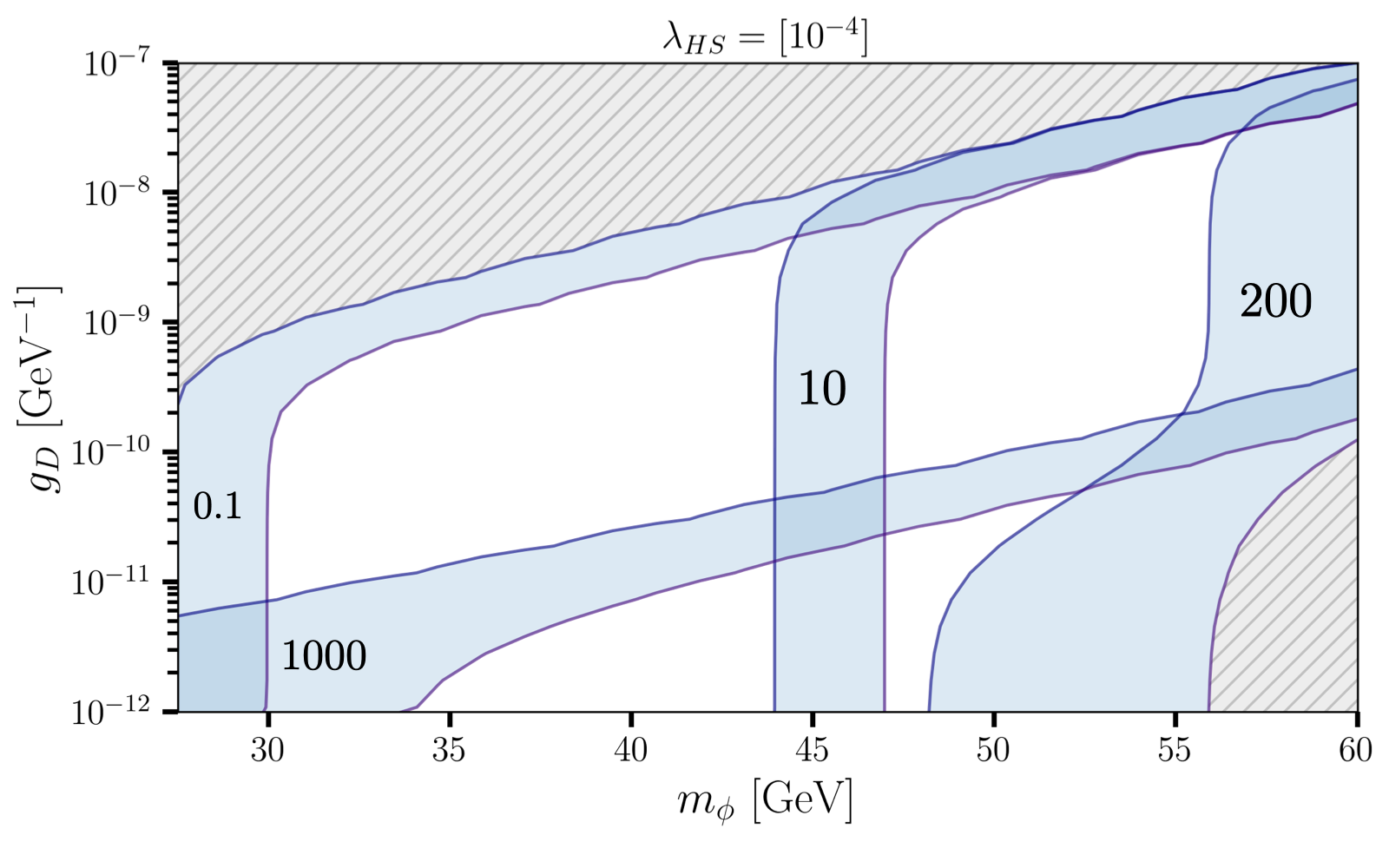}
\caption{Dark coupling $g_D$ vs the pseudo-scalar mass $\mphi$ for $T_{\rm RH}=3$~GeV with levels for different range values of $\lambda_{ HS}$ indicated in units of $10^{-4}$. Each band shows the correct parameter space for dark photon dark matter within the masses $m_{\gamma'}\in [1, 5]$~GeV, marked in blue and purple, respectively. The white area between the lines can be accessed for those dark photon masses by choosing a value of $\lambda_{HS}$ between the levels. All cosmological restrictions have been applied. The hatched areas are not viable, either because they cannot reproduce the correct relic abundance, or because they do not satisfy cosmological constraints.}
\label{fig:parameter_space_cosmo}
\end{figure}

\subsubsection{Viable dark matter parameter space }
\label{sec:viableDM}
Figure~\ref{fig:parameter_space_cosmo} displays the viable parameter space in the $g_D-m_{\phi}$ plane, where a dark photon with mass in our target range can account for the total observed dark matter abundance, $\Omega_{\gamma'} h^2 \sim 0.12$, fulfilling all cosmological restrictions. Each light blue band corresponds to an isocontour of constant $\lambda_{HS}$, with the dark photon mass varying from $m_{\gamma'} = 1$~GeV (blue) to $m_{\gamma'} = 5$~GeV (purple). The unshaded (white) regions to the right can be populated by increasing $\lambda_{HS}$ beyond the values shown on the left.

The behavior at (small) $\lambda_{HS}< 10^{-2}$ is divided in two regimes:  the vertical contours indicate that the relic abundance is largely independent of $g_D$. This corresponds to a scenario in which the mediator $\phi$ is produced via freeze-in through the Higgs portal, with dark matter subsequently generated through the decay $\phi \to \gamma\gamma'$. The right upper part of the plot suggests a convergence of contours with different values of $\lambda_{HS}$ as $g_D$ increases, pointing to an independence of the Higgs portal coupling. This is indeed what it is expected at larger $\mphi$, where number density suppression becomes significant, allowing $g_D$ to increase and dominate DM production through the process $f\bar{f} \rightarrow \phi\gamma'$, with only a subdominant contribution from the Higgs portal interaction.
Interestingly, a second overlap region emerges at lower values of $g_D$, for example around $g_D \sim 10^{-11}$~GeV$^{-1}$ and $\mphi \sim 47$~GeV, where the contours corresponding to $\lambda_{HS} = 10^{-3}$ and $0.1$ intersect. This suggests that within this region of the $g_D-\mphi$ plane,  two distinct values of $\lambda_{HS}$ can produce the correct relic abundance. This can be understood because they represent two different thermal histories for $\phi$. For smaller values of the Higgs coupling, the pseudo-scalar never thermalizes and its abundance is set by freeze-in. In contrast, a higher value of the coupling leads to scalar thermalization (see figure~\ref{fig:3}). That is, the parametric relationship between the coupling and the abundance changes. For the latter scenario, the dark photon falls in the category of a super-WIMP.

\section{Collider phenomenology} \label{sec:pheno}

We present here the existing constraints given by the LHC results on non-pointing photon searches, and compare them with the bounds coming from data on Higgs bosons production and its decays to invisible or undetected final states. Furthermore, we discuss how they impact the viability of the DM model parameter space.

\subsection{LHC constraints}

\subsubsection{Bound from non-pointing photons search}

The analysis motivating this work uses the full Run 2 data set of proton-proton collisions delivered by the LHC at a center-of-mass energy of $\sqrt{s}=13$~TeV between 2015 and 2018 and recorded by the ATLAS detector, corresponding to an integrated luminosity of $139~{\rm fb}^{-1}$.
The experiment searches for delayed and non-pointing photons, which would originate from the displaced decay of a neutral long-lived particle (LLP)~\cite{ATLAS:2022vhr}. The measurement is triggered by a prompt electron or muon coming from associated production with the Higgs.\footnote{We note that Higgs production via gluon fusion could also be taken into account for this search by using a single-jet trigger instead, requiring $p^j_T>100-200$~GeV~\cite{ATLAS:2016wtr}. We have estimated that, at leading order, the corresponding cross-section is of an order of magnitude similar to those for production with an associated lepton.} The results are interpreted in a scenario in which the LLPs are produced in pairs in exotic decays of the $125$~GeV Higgs boson, with each LLP subsequently decaying into a photon and a particle that escapes direct detection, giving rise to missing transverse momentum. No significant excess is observed above the expectation due to background processes, so upper limits are set on the branching ratio of the exotic decay of the Higgs boson. In addition, a model-independent limit is also provided for the production cross-section of photons with large values of displacement and time delay.

\begin{figure}[t!]
\centering
\includegraphics[width=0.65\textwidth]{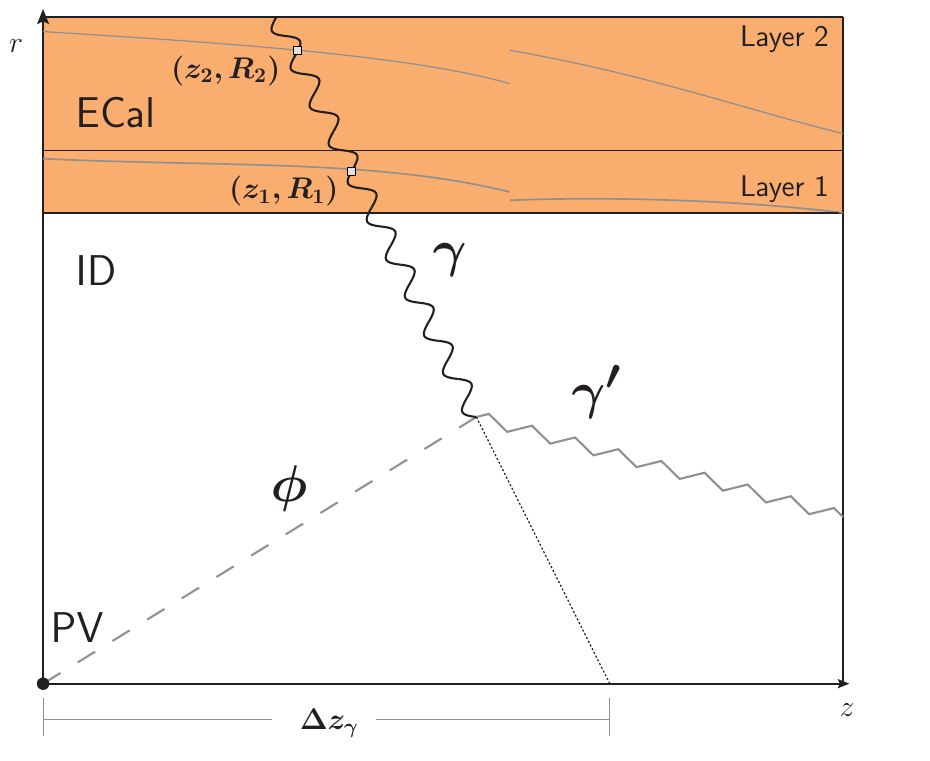}
\caption{A schematic for the pointing $\Delta z_\gamma$ measurements used in the analysis, with the photon energy deposits shown in the longitudinal layers of the LAr EM calorimeter. The $R_i$ represent the radial depths, that is, the locations where the dominant energy depositions occur within the ECal (see appendix~\ref{sec:appCollider}).}
\label{fig:calo_diagram}
\end{figure}
The displacement and time delay of the photon are both reconstructed through the electromagnetic shower of the photon within the ATLAS Electromagnetic Calorimeter (ECal). The time delay $t_\gamma$ is defined as the relative arrival time to the ECal, with respect to what is expected from a prompt photon. Furthermore, the non-pointing parameter $|\Delta z_\gamma|$ is defined as the separation between the extrapolated origin of the photon and the position of the primary vertex of the event, projected onto the beamline in the $z-r$ plane, as depicted in figure~\ref{fig:calo_diagram}. Both variables are determined in the first two layers of the ECal, with the $|\Delta z_\gamma|$ measurement also being possible due to the varying lateral segmentation of the detector, which allows reconstruction of the flight direction of the photon. Then, for prompt decays, both $|\Delta z_\gamma|$ and $t_\gamma$ are naively expected to be zero, meaning that these are useful handles for identifying neutral long-lived particles decaying into photons.

Let us now briefly describe the event selection procedure in the experimental analysis.\footnote{A more detailed description of the recast can be found in~\cite{Delgado:2022fea,Duarte:2023tdw}. The technical details of the current refinements on the simulation of our signal events, their analysis and statistical interpretation are described in appendix~\ref{sec:appCollider}.} The photons and charged leptons had to satisfy selection cuts, as described in~\cite{Duarte:2023tdw}. In the search, in addition to the lepton trigger that requires a prompt electron or muon with $p_T>27$~GeV, the selected events must have at least one photon candidate, being divided into two orthogonal final states, with the $1\gamma$ final state requiring exactly one photon (single-photon channel) and the $\geq 2 \gamma$ final state requiring two or more photons (multiphoton channel). Regardless of this classification, the analysis uses the information $t_\gamma$ and $|\Delta z_\gamma|$ from the photon with the largest $p_{\gamma T}$ in the barrel, on which a cut $E_{\rm cell}>10$~GeV is applied ($E_{\rm cell}$ is defined as the maximum energy deposit of the shower in the middle layer of the ECal). Events are also required to have a minimum value of missing transverse energy ${\rm MET} > 50$~GeV to account for the escaping particles, in our case, the dark photons and neutrinos from leptonic decays of the associated system.

In order to contribute to any of the aforementioned channels, the analyzed photon must satisfy $|\Delta z_\gamma|\,>300$~mm. Moreover, for the single-photon channel, the search requires $1.5<t_\gamma<12$~ns, while the multi-photon channel needs $1<t_\gamma<12$~ns. The number of events measured in these bins, as well as the background, is given in table II of~\cite{ATLAS:2022vhr}.

\begin{table}[tb]
\centering
\begin{tabularx}{\textwidth}{|c| >{\centering\arraybackslash}X |
                                    >{\centering\arraybackslash}X |
                                    >{\centering\arraybackslash}X |
                                    >{\centering\arraybackslash}X |}
\hline
Cut & \multicolumn{2}{c|}{$c\,\tau_\phi=0.69$~m} & \multicolumn{2}{c|}{$c\,\tau_\phi=4.4$~m} \\
\hline
Trigger, $p_{\gamma T}>10~{\rm GeV}$, Acceptance & \multicolumn{2}{c|}{71} & \multicolumn{2}{c|}{44} \\
Isolation, Efficiencies, Z-veto & \multicolumn{2}{c|}{39} & \multicolumn{2}{c|}{18} \\
\hline
& \multicolumn{4}{c|}{Channel} \\
& $1$ & $2+$ & $1$ & $2+$ \\
\hline
$E_{\rm cell}>10$~GeV & 15 & 20 & 11 & 3.2 \\
${\rm MET} > 50$~GeV & 6.2 & 8.3 & 5.4 & 1.3 \\ 
$|\Delta z_\gamma|>300$~mm & 2.2 & 1.8 & 2.6 & 0.46 \\
$t_\gamma > 1\, (1.5)$~ns & 0.20 & 0.45 & 0.89 & 0.27 \\
\hline
\end{tabularx}
\caption{Cutflow showing efficiency (in $\%$) for the displaced photon search, taking $m_\phi=60$~GeV, $m_{\gamma'}=1$~GeV. Here, we combine $Zh$, $W^\pm h$ and $t\bar th$ production processes. We show results for $g_D=5.7\times10^{-10}$ and $1.6\times10^{-10}$~GeV$^{-1}$, which respectively imply decay lengths $c\,\tau_\phi=0.69$ and $4.4$~m. We distinguish between single ($1$) and multi- ($2+$) photon channels.}
\label{tab:cutflow}
\end{table}
In order to understand the impact of each cut of the displaced photon analysis, we show the cutflow for two benchmark points on table~\ref{tab:cutflow}. The two points are chosen in order to characterize scenarios with relatively short and long $\phi$ decay lengths, $c\,\tau_\phi=0.69$ and $4.4$~m, respectively. The table shows the impact due to selection cuts before and after the events are assigned to single ($1$) or multi- ($2+$) photon channels. Before this assignment, the efficiency for the point with a short decay duration is around $40\%$, while the one with a long decay duration drops below $20\%$. This is reasonable, as LLPs with a long lifetime are more likely to escape the detector and not pass the acceptance cut.

After assignment to the two channels, table~\ref{tab:cutflow} shows a dramatic reduction in signal efficiency in both benchmarks. After cuts, the efficiency for the benchmark with a short decay length drops to about $0.65\%$, while that for the long decay length is reduced to about $1.2\%$. In fact, even though the first benchmark is the least affected by the acceptance, its short lifetime leads to events with small values of $t_\gamma$ and $|\Delta z_\gamma|$, which have a hard time passing the cuts. In contrast, the second benchmark with a long decay length, which is strongly constrained by acceptance, usually has higher values of $t_\gamma$ and $|\Delta z_\gamma|$.

The cutflow also hints that the single-photon channel will have a larger efficiency in scenarios with a long decay length, while the multi-photon channel is more efficient for the shorter decay-length cases. Indeed, if the LLP has a long lifetime, it is more likely for one of its daughter photons to be produced outside the detector, leading to a much larger efficiency for the single-photon channel. A broader scan finds that this is the trend. However, in general, the difference in efficiency is not dramatic. 

As summarized in appendix~\ref{sec:statistical}, the number of simulated events that pass the cuts performed in the ATLAS search provides a bound on the maximum possible value of the branching ratio ${\rm BR}(h\to \phi \phi)$, given by eq.(\ref{eq:events}).\footnote{In the following, we will focus on the single and multi-photon channel ``combination'' number of surviving events.} In turn, this bound can be translated to a limit on the maximum possible value of the $\lambda_{HS}$ coupling allowed by the experiment, using eq.~(\ref{eq:GammaHiggs}), for a certain parameter space point $(m_{\gamma'},\, m_\phi,\, g_D)$ in the target region we proposed in section \ref{sec:parameterspace}.

\subsubsection{Bound from Higgs invisible decays}

Higgs invisible decays refer to situations where the Higgs boson decays either to stable neutral particles or to very long-lived neutral particles which decay outside the detector. Such decays are reflected on $B_{\rm inv}$, the branching ratio of the Higgs into the invisible final state. The strongest limit in the literature to this branching ratio comes from a combination of direct searches, performed by ATLAS~\cite{ATLAS:2023tkt}, reading $B_{\rm inv}<0.107$.

In our scenario, the limit on $B_{\rm inv}$ would apply to regions of the parameter space where $\phi$ is expected to decay outside the detector. If we assume a spherical detector of radius $L_{\rm det}$, the probability for $\phi$ decaying anywhere outside the detector is:
\begin{equation}
\label{eq:probDecay}
P(L_{\rm det},\,\infty)=\exp[-\frac{L_{\rm det}}{\gamma_{\rm rel}\,\beta_{\rm rel}\,c\,\tau_\phi}]~.
\end{equation}
where $\gamma_{\rm rel}=E_\phi/m_\phi$ and $\beta_{\rm rel}=|\vec p_\phi|/E_\phi$. Eqs.~(\ref{eq:GammaHiggs}) and~(\ref{eq:probDecay}) can then be used to estimate the probability of a long-lived $\phi$ contributing to the Higgs invisible decay channel. Namely, one needs to combine the $h\to\phi\phi$ branching ratio with the probability that both scalars decay outside the relevant detector. Thus:
\begin{equation}
 B_{\rm inv}={\rm BR}(h\to\phi\phi)\exp[-\frac{2L_{\rm det}}{\gamma_{\rm rel}\,\beta_{\rm rel}\,c\,\tau_\phi}]<0.107
\end{equation}
In order to estimate $\gamma_{\rm rel}\,\beta_{\rm rel}$, we assume that the Higgs decays at rest, so $E_\phi=m_h/2$. Then:
\begin{equation}
\label{eq:gammabetaEstimate}
 \gamma_{\rm rel}\,\beta_{\rm rel}=\sqrt{\frac{m_h^2}{4m_\phi^2}-1}
\end{equation}
Thus, we have $\gamma_{\rm rel}\,\beta_{\rm rel}$ between $6.2 - 0.29$ for $m_\phi$ between $10 - 60$~GeV, respectively. These can be interpreted as the minimum values for $\gamma_{\rm rel}\,\beta_{\rm rel}$ when the Higgs is boosted.

In the following, we shall compare our bounds from non-pointing photon searches with a conservative limit from Higgs invisible decays. For this specific constraint, we will indeed assume that the Higgs decays at rest and that the ATLAS ECal is spherical, with $L_{\rm det}=1.97$~m~\cite{ATLAS:2008xda}. This implies:
\begin{equation}\label{eq:BinvBound}
 {\rm BR}(h\to\phi\phi)<0.107\exp[\frac{g_D^2\, c_W^2\,(10^{-16}~{\rm GeV}^{-1})}{8\pi}\frac{m_\phi^4}{\sqrt{m_h^2-4m_\phi^2}}\left(1 - \frac{m_{\gamma'}^2}{m_\phi^2}\right)^3]
\end{equation}
This bound can then be applied directly to each parameter space point for the values of each parameter space point $(m_{\gamma'},\, m_\phi,\, g_D)$, giving a maximum allowed value for $\lambda_{HS}$ using eq.~(\ref{eq:GammaHiggs}).

\subsubsection{Bound from a fit to combined measurements}

The CMS~\cite{CMS:2018uag} and ATLAS~\cite{ATLAS:2019nkf,ATLAS:2022vkf} collaborations have performed a likelihood fit of a combination of their data on Higgs production and decay, in order to constrain the Higgs couplings and its branching ratios into invisible and undetected particles.

In our case, since $\phi$ does not mix with the Higgs, the only deviation with respect to the SM Higgs properties comes from exotic decays. In particular, since displaced photon searches were not taken into account in the fit, our scenario is subject to the bound on the Higgs branching ratio to undetected particles $B_{\rm und}<0.12$~\cite{ATLAS:2022vkf}. This bound is also reinterpreted, for each parameter space point $(m_{\gamma'}, m_\phi)$ using eq.~(\ref{eq:GammaHiggs}), as a bound on $\lambda_{HS}$. Note that it is independent of the value of the $g_{D}$ coupling.

\subsection{Collider results and consequences on dark matter}

\begin{figure}[t!]
\centering
\includegraphics[width=0.48\textwidth]{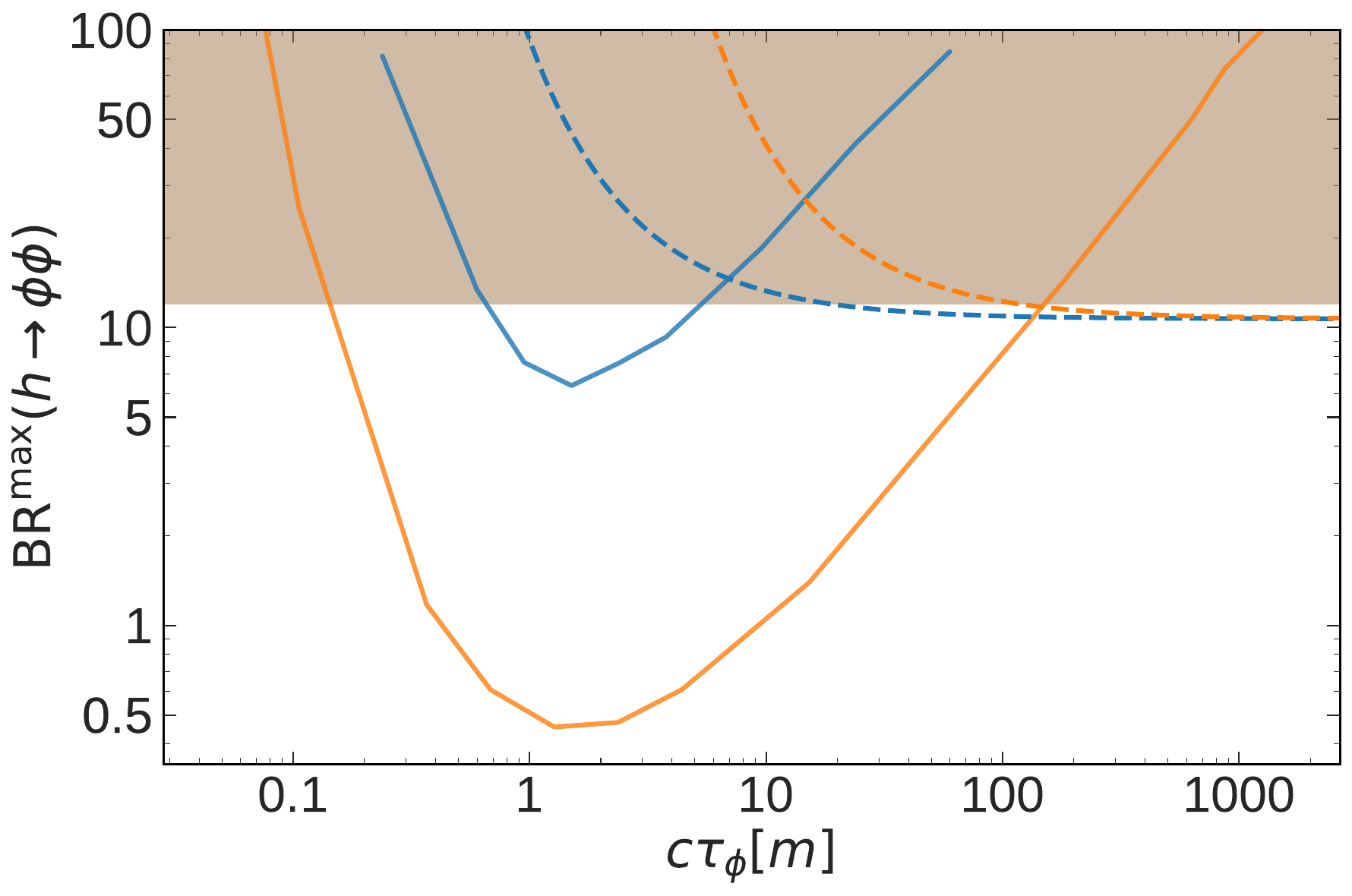}
\hfill
\includegraphics[width=0.48\textwidth]{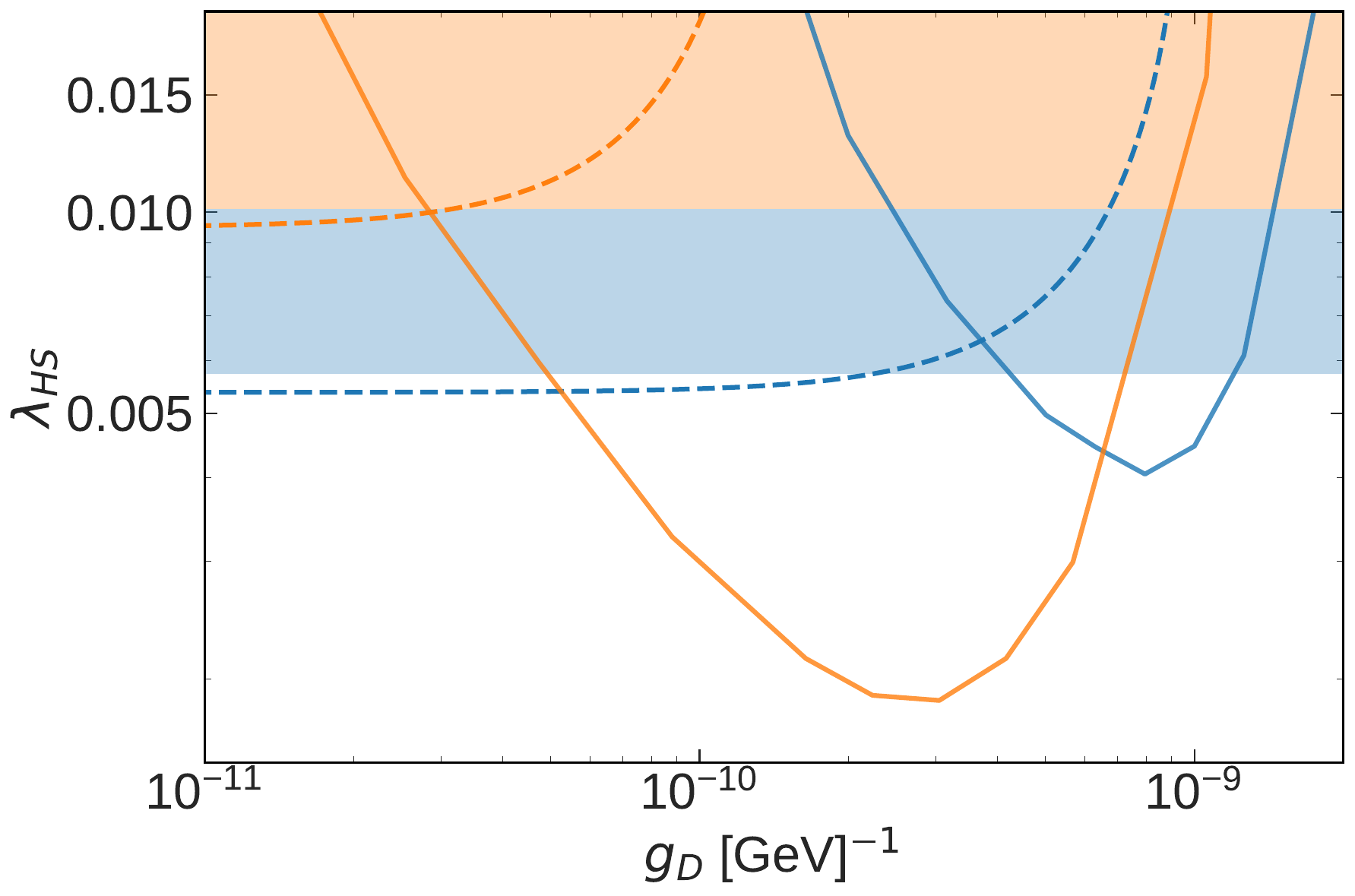}
\caption{Left: Constraints of ${\rm BR}(h\to\phi\phi)$ as a function of $c\,\tau_\phi$. Right: Constraints of $\lambda_{HS}$ as a function of $g_D$. In both panels, we present bounds coming from displaced photons (solid), Higgs invisible decays (dashed) and Higgs decays to undetected particles (shaded regions). Benchmarks points with $m_\phi=30,\,60$~GeV are shown in blue and orange, respectively, both having $m_{\gamma'}=1$~GeV.}% 
\label{fig:ColliderSlices}
\end{figure}
We present constraints to the model from the LHC experiments in figure~\ref{fig:ColliderSlices}. In the left panel, we show the bounds on the $h\to\phi\phi$ branching ratio as a function of the decay length $c\,\tau_\phi$. Constraints coming from the displaced photon search and invisible Higgs decays are shown in solid and dashed lines, respectively. The curves in orange (blue) correspond to $m_\phi=60$~GeV (30~GeV), with $m_{\gamma'}=1$~GeV. Moreover, the shaded region displays the bound from undetected Higgs decays. The maximum values allowed for the branching ratio are those below each curve for a given $c\,\tau_\phi$.

For both benchmark $\phi$ masses, we find that the displaced-photon search has the highest sensitivity for $c\,\tau_\phi\approx2$~m. For this decay length, branching ratios down to about $0.5\%$ can be probed for $m_\phi=60$~GeV. This sensitivity is lower for $m_\phi=30$~GeV, being able to constrain branching ratios around $7\%$. The increased sensitivity for larger masses was explained in~\cite{Duarte:2023tdw}. On the one hand, larger masses lead to slower moving LLPs, such that the displaced photon arrives at the ECal with a larger delay. On the other hand, with a smaller boost the momenta of the LLP decay products are much less collimated, so the displaced photon points less towards the primary vertex, leading to a larger $|\Delta z_\gamma|$. Thus, LLPs with larger masses have a greater probability that their displaced photons will be assigned to the signal region of the search.

The left panel of figure~\ref{fig:ColliderSlices} also shows constraints from invisible Higgs decays. For sufficiently large decay lengths, these bound the branching ratio to $10.7\%$. As the decay length becomes shorter, the bound weakens as the probability of both LLPs to decay outside the detector drops. Nevertheless, we find that for $m_\phi=60$~GeV (30~GeV), these are the strongest bounds on the branching ratio as long as $c\,\tau_\phi\gtrsim100$~m (20~m). In addition, the figure also shows the flat limit on $12\%$ coming from Higgs decays to undetected final states. The latter are particularly relevant for short $\phi$ decay lengths, where both the displaced photons search and the Higgs invisible decay measurements lose their sensitivity.

In the right panel of figure~\ref{fig:ColliderSlices} we present the same information in terms of $\lambda_{HS}$ and $g_D$. Given $m_\phi$ between 30 and 60~GeV, the displaced photon search applies bounds on $\lambda_{HS}$ for values of $g_D$ between $\mathcal O(10^{-11})-\mathcal O(10^{-9})$~GeV$^{-1}$, consistent with our expectations in figure~\ref{fig:parameter_space}. For $m_\phi=60$~GeV, the strongest bound is $\lambda_{HS}\lesssim2\times10^{-3}$, for $g_D\sim3\times10^{-10}$~GeV$^{-1}$. Furthermore, if $m_\phi=30$~GeV, the strongest bound is $\lambda_{HS}\lesssim4\times10^{-3}$ instead, valid for $g_D\sim8\times10^{-10}$~GeV$^{-1}$. Searches for Higgs decay into invisible or undetected states also enforce constraints on $\lambda_{HS}\lesssim10^{-2}$ ($\lambda_{HS}\lesssim5\times10^{-3}$) for $m_\phi=60$~GeV (30~GeV).

\begin{figure}[t!]
\centering
\includegraphics[width=0.48\textwidth]{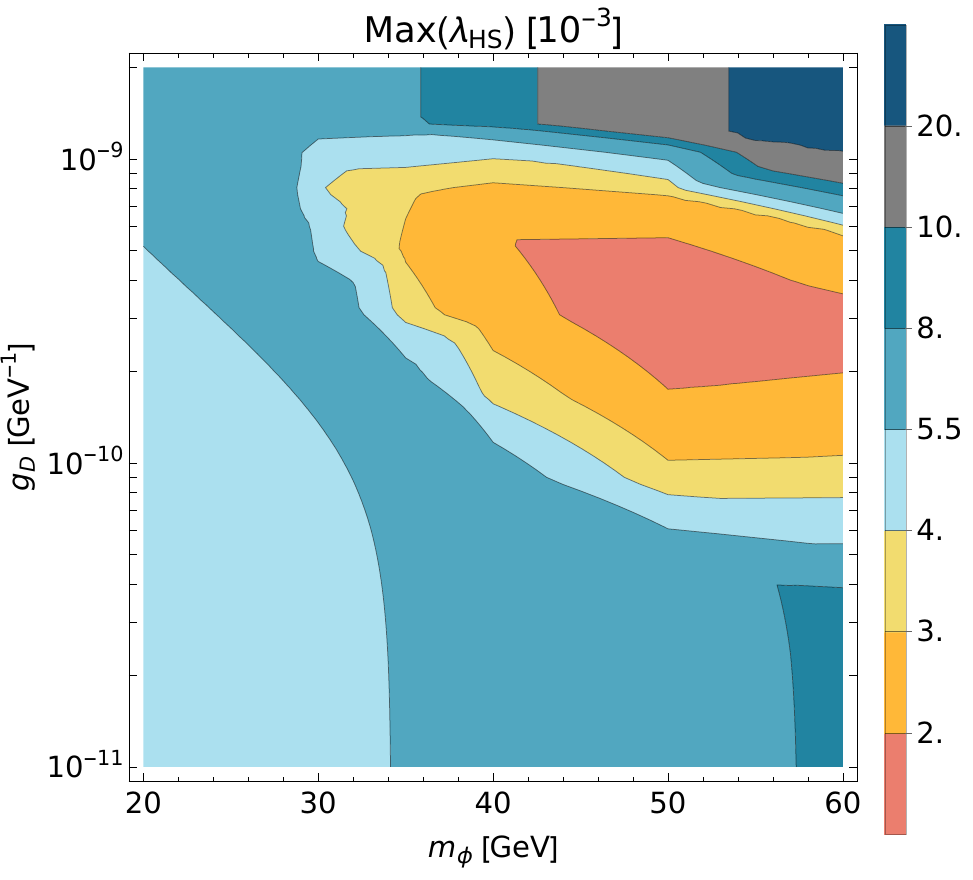}
\hfill
\includegraphics[width=0.48\textwidth]{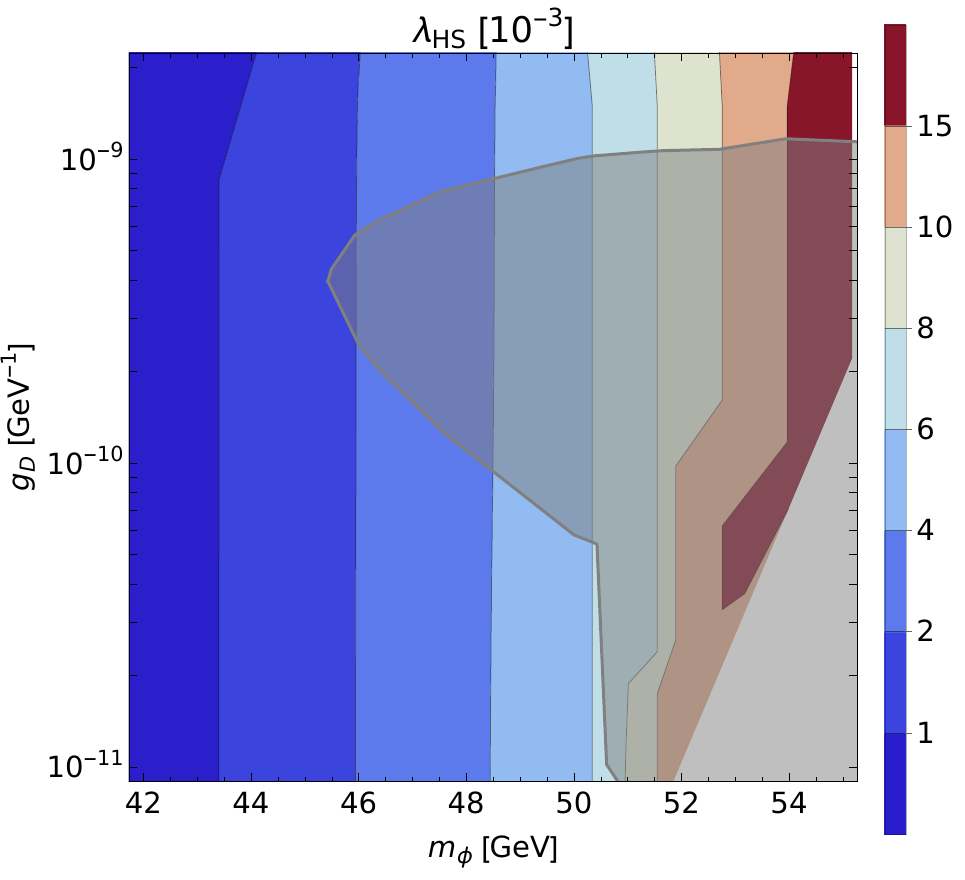}
\caption{Left: Exclusion regions from the collider analysis, considering $m_{\gamma'} = 1$ GeV. Each color represents the maximum value that $\lambda_{HS}$ may acquire. Right: Contours of $\lambda_{HS}$ in the $g_D$–$\mphi$ plane corresponding to the parameter space where the dark photon constitutes a viable dark matter candidate, under the assumption that the mediator never thermalizes. Regions excluded by collider constraints are shown in gray. In this case we keep $m_{\gamma'} = 1$ GeV and set $T_{\rm RH} = 3$ GeV.}
\label{fig:ColliderContour}
\end{figure}
We present our final result on the left panel of figure~\ref{fig:ColliderContour}, as contour lines giving the maximum allowed $\lambda_{HS}$ in the $g_D - m_\phi$ space, for $m_{\gamma'}= 1$~GeV. The contours in the middle right zone of the plot, with colors from red to light blue, constrain $\lambda_{HS}$ to values below $2$ -- $5.5\times 10^{-3}$, and come from the search for displaced photons. On the other hand, the vertical lines on the upper right, with colors in blue, gray and dark-cyan, stem from the Higgs to undetected bound and, as we mentioned earlier, are independent of the value of $g_D$. The rest of the contours, on the left and lower parts of the plot, are governed by the Higgs to invisible bound, following eq.~(\ref{eq:BinvBound}). These contours are consistent with our results in figure~\ref{fig:ColliderSlices}, and demonstrate the synergies of the three different analyses when constraining the parameter space.

Still in the left panel of figure~\ref{fig:ColliderContour}, it is important to note that the bounds presented by the vertical lines at the bottom of the plot, due to Higgs to invisible constraints, can be extrapolated to arbitrarily small values of $g_D$. The reason is that smaller values of $g_D$ will only increase the decay length of $\phi$, which is already expected to escape undetected for $g_D\sim\mathcal O(10^{-11})$~GeV$^{-1}$. The vertical lines at the top, in contrast, cannot be extrapolated to much larger values of $g_D$, since in that case the scalars will decay promptly and will not contribute to the Higgs to undetected channel. Finally, the bound of $\lambda_{HS}\lesssim5\times10^{-3}$ applied on the left part of the plot, again coming from Higgs to invisible, can be extrapolated to arbitrarily small $\phi$ masses. This implies that for $m_\phi<m_h/2$ and $g_D\lesssim10^{-9}$~GeV$^{-1}$, collider searches constrain $\lambda_{HS}$ to values under $\mathcal O(10^{-3})$.

The conclusion above has important consequences for our DM results. Recall from figure~\ref{fig:parameter_space_cosmo}, as well as the discussion in section~\ref{sec:viableDM}, that $\phi$ could thermalize in the early Universe, in some regions of parameter space, for values of $\lambda_{HS}\sim\mathcal O(0.1)$. What our collider results show is that, given our assumption on $T_{\rm RH}=3$~GeV and $m_{\gamma'}=1$~GeV, we can rule out the possibility of having a thermal $\phi$ after reheating and a super-WIMP dark matter candidate.

In addition, collider searches can also constrain the parameter space where both dark sector particles are produced only by freeze-in processes, displayed as vertical lines in figure~\ref{fig:parameter_space_cosmo}. That scenario is presented in the right panel of figure~\ref{fig:ColliderContour}, where we show the contours of $\lambda_{HS}$ compatible with dark photon DM with $m_{\gamma'} = 1$ GeV, $\lambda_{HS} < 2\times10^{-2}$  and $T_{\rm RH} = 3$ GeV. In the plot, the gray region is ruled out by a combination of displaced photon and Higgs to invisible searches, and is determined by the intersection of the relic density contours and the constraints enforced by the left panel. Thus, we conclude that collider searches can indeed give valuable constraints to DM models such as ours.

Let us briefly comment on how we expect our constraints to change when varying $m_{\gamma'}$ and $T_{\rm RH}$. From the collider side, the bounds remain valid for negligible values of $m_{\gamma'}$. However, higher values, approaching $m_{\phi}$, can provide an additional phase space suppression to the $\phi$ decay width, leading to a larger lifetime (see eq.~(\ref{eq:lifetime})). 
This effect could shift our exclusion limits, allowing us to probe larger values of $g_D$. However, increasing the dark photon mass would also reduce the transverse momentum of its associated SM photon, causing fewer photons to pass the energy threshold in the non-pointing photon search and thus decreasing the sensitivity. It is worth noting that all collider results remain completely independent of the reheating temperature $T_{\rm RH}$.

From the perspective of DM production and the cosmological history of the dark sector particles, shifting the DM mass to lower values --- while keeping all other parameters fixed --- requires reducing the Boltzmann suppression of the mediator in order to reproduce the correct relic density, correspondingly decreasing its mass. This effect can potentially tighten structure formation constraints, as the mediator becomes longer-lived and the dark matter lighter. However, these effects can be partially mitigated by lowering the reheating temperature. Additionally, increasing the dark matter mass while respecting $\mphi < m_h/2$, also requires a lower reheating temperature. Finally, increasing $T_{\rm RH}$ can easily lead to dark matter overproduction.  Firstly, due to UV freeze-in through the dark photon portal, and secondly, due to the opening of additional production channels for the mediator, along with reduced Boltzmann suppression. Therefore, a thermalization of the mediator is expected for smaller values of $\lambda_{HS}$. Within the region of parameter space probed by collider searches, the conditions for obtaining a viable dark photon DM scenario are optimally satisfied for a reheating temperature around $T_{\rm RH} \sim 3$~GeV. This identifies a well-defined target for the model that can be tested experimentally.

\section{Conclusions}\label{sec:conc}

Freeze-in of DM with a low reheating temperature has gained significant attention in recent years. This framework allows for larger coupling constants, making specific realizations of these scenarios potentially testable in various types of laboratory experiments, e.g. direct detection and colliders. In this work, we have been interested in the possibility of testing such a scenario with LHC searches for displaced photons. These photons arise from the disintegration of LLPs, which in turn are produced from exotic Higgs decays. Our aim was to identify a simple model capable of reproducing the correct DM relic density, satisfying cosmological constraints, and leaving a signal at colliders.

Our model features dark photon $\gamma'$ dark matter with a pseudo-scalar mediator $\phi$, both odd under a $\mathbb{Z}_2$ symmetry. These particles interact with the SM through a Higgs portal, in eq.~(\ref{lag1}), and a dark photon portal, eq.~(\ref{eq:portal}). From the cosmological perspective, we identify a consistent DM scenario in which the dark photon accounts for the relic abundance via freeze-in, through decays of the non-thermal $\phi$ produced via the Higgs portal. The viability of this mechanism is sensitive to the reheating temperature, with a golden window around $T_{\rm RH}\sim 3$~GeV allowing the correct abundance while avoiding overproduction. Moreover, we also identify a region of parameter space where the mediator thermalizes with the Standard Model plasma and, after freezing out, significantly impacts the final relic abundance of the dark photon.
 
We have shown that LHC searches for displaced photons, as well as searches for invisible Higgs decays, can significantly constrain the parameter space of our dark photon DM model. In particular, we find that displaced photon searches provide the most stringent bounds for mediator decay lengths of around a few meters, reaching sensitivities to Higgs branching ratios to the pseudo-scalars as low as 0.5\%. Our results imply constraints on the portal coupling $\lambda_{HS}$ of $\mathcal{O}(10^{-3})$, which completely rules out the possibility having a thermal $\phi$ in the early Universe, and reinforce a purely non-thermal picture. The latter evades collider bounds for light enough mediator masses, being subject to constraints only for masses above $\sim45$~GeV.

To summarize, our results identify a viable region of parameter space in a DM model featuring a long-lived pseudo-scalar mediator, whose phenomenology can be tested via displaced photon signatures at the LHC. Despite their potential to probe suppressed interactions, such signals remain largely unexplored in the context of DM models. The present analysis illustrates how this channel can be exploited to access otherwise elusive scenarios. We hope that these findings support the case for the development of dedicated experimental searches targeting displaced photon signatures arising from well-motivated dark sector models.
    
\section{Acknowledgments}
We thank Sasha Pukhov for his constant help with {\tt micrOMEGAs}. J.J.P., W.R.~and D.Z.~acknowledge financial support from the \textit{Dirección de Fomento de la Investigación} at PUCP, though grant DFI-PUCP-PI1144. B.D.S. was funded by ANID FONDECYT postdoctorado 2022 No.~3220566 and ANID FONDECYT Iniciación No.~1125181. B.D.S. also acknowledges support from ANID — Millennium Science Initiative Program ICN2019_044 . P.A. acknowledges support from FONDECYT project Nº 1251613. L.D. acknowledges support from PEDECIBA (Uruguay).

\appendix

\section{Non-relativistic average cross sections}\label{app_a}

In the following, we compute the average annihilation cross section times velocity for $\phi\gamma' \rightarrow f\bar{f}$, with $f$ any SM fermion, assuming that the coannihilation of the initial states is non-relativistic. In each case, we have replaced $\ev{v^2}  \rightarrow 6T/\mu$, with $\mu = m_\phi m_{\gamma'}/(m_\phi  + m_{\gamma'})$.

In the case of SM charged leptons in the final state and neglecting the fermion mass, 
\begin{eqnarray}
    \ev{\sigma v}_{\phi\gamma' \rightarrow l\bar{l}} = \frac{e^2 g_D^2 \left[5 (m_\phi + m_{\gamma'})^4 - 12c_W^2(m_\phi + m_{\gamma'})^2m_Z^2 + 8c_W^4 m_Z^4\right]}{24 \pi c_W^2 \sqrt{m_\phi m_{\gamma'}}\left[(m_\phi + m_{\gamma'})^2 - m_Z^2\right]^2} T\,.
\end{eqnarray}
In the case of SM neutrinos, the mediator is only the $Z$ boson, and we obtain
\begin{eqnarray}
    \ev{\sigma v}_{\phi\gamma' \rightarrow \nu\bar{\nu}} = \frac{ e^2 g_D^2 (m_\phi + m_{\gamma'})^4 }{24 \pi c_W^2 \sqrt{m_\phi m_{\gamma'}} ((m_\phi + m_{\gamma'})^2 - m_Z^2)^2} T
\end{eqnarray}
For up-type quarks we have
\begin{eqnarray}
    \ev{\sigma v}_{\phi\gamma' \rightarrow u\bar{u}} = \frac{e^2 g_D^2 \left[17 (m_\phi + m_{\gamma'})^4 - 40c_W^2(m_\phi + m_{\gamma'})^2 m_Z^2 + 32c_W^4 m_Z^4\right]}{72 \pi c_W^2 \sqrt{m_\phi m_{\gamma'}} \left[(m_\phi + m_{\gamma'})^2 - m_Z^2\right]^2} T,
\end{eqnarray}
and for down-type quarks
\begin{eqnarray}
    \ev{\sigma v}_{\phi\gamma' \rightarrow d\bar{d}} = \frac{e^2 g_D^2 \left[5 (m_\phi + m_{\gamma'})^4 - 4c_W^2(m_\phi + m_{\gamma'})^2 m_Z^2 + 8c_W^4   m_Z^4\right]}{72 \pi c_W^2 \sqrt{m_\phi m_{\gamma'}}\left[(m_\phi + m_{\gamma'})^2 - m_Z^2\right]^2} T.
\end{eqnarray}

\section{Displaced photons recast analysis improvements}\label{sec:appCollider}

\subsection{LHC signal simulation and improved non-pointing variable}

In our signal simulation, we consider that $\phi$ is pair-produced promptly in proton-proton collisions as decay products of on-shell Higgs bosons $h$. Then, each long-lived scalar decays to one dark photon $\gamma'$ and a photon $\gamma$ that can be detected at the ECal. These photons, coming from an LLP, can lead to a non-pointing photon signal. We considered the triggering leptons coming from $p\,p\to W^\pm h$, $p\,p\to Z\,h$ and $p\,p\to t\,\bar t\, h$ processes. The model was built in~\texttt{FeynRules 2.3.43}~\cite{Christensen:2008py,Alloul:2013bka}, and events were generated with \texttt{MadGraph5\_aMC@NLO 2.9.7}~\cite{Alwall:2014hca}, which uses \texttt{LHAPDF6}~\cite{Buckley:2014ana}. \texttt{PYTHIA 8.244}~\cite{Sjostrand:2006za} performs the $\phi$ scalar decay, parton showering and hadronization.

The interested reader can find details of how $t_\gamma$ was modeled in~\cite{Delgado:2022fea,Duarte:2023tdw}. However, for this work we carry out a refined calculation of $|\Delta z_\gamma|$. The non-pointing variable used in~\cite{ATLAS:2022vhr} is defined as~\cite{Nikiforou:2014cka,Mahon:2021bai}:
\begin{equation}
  \label{eq:deltaZexp}
  \Delta z_\gamma = \frac{z_1 R_2- z_2 R_1}{R_2 - R_1}.
\end{equation}
where $z_i,\,R_i$ are the coordinates of the position of the centroid of the  electromagnetic shower associated to each photon on the first two layers of the detector, as illustrated in figure~\ref{fig:calo_diagram}. The dependence of the radial depths $R_i$ on the reconstructed pseudorapidity are shown on page 48 (figure 5) in~\cite{ATLAS:2009zsq}.

For this work we improve the calculation of $|\Delta z_\gamma|$ by using the information on $R_i$ found in~\cite{ATLAS:2009zsq}. First, we use the truth-level vectors $\vec r_\phi$ and $\vec p_\gamma$, for the $\phi$ decay position and momentum of its daughter photon, respectively, to describe the photon's trajectory inside the detector:
\begin{equation}
\label{eq:L2}
 \vec L=\vec r_{\phi}+\alpha\,\frac{\vec p_\gamma}{|\vec p_\gamma|}
\end{equation}
where $\alpha$ is an evolution parameter. Using this equation, one can calculate the $z_1$ and $z_2$ coordinates where $\vec L$ intersects $r=R_1$ and $r=R_2$ surfaces, for specific $R_i$, regardless of the azimuthal angle $\varphi$. In order to determine which value of $R_i$ we should use, we first find the pseudorapidity $\eta_i$ of the photon as it enters each layer, at $r=1500$ and $1590$~mm, respectively (see figure~5.4 of~\cite{ATLAS:2008xda}). Notice that this pseudorapidity refers to a region of the detector, as measured from the interaction point. So, for each value of $\eta_i$, we find the appropriate $R_i$ from the corresponding curves of figure~5 of~\cite{ATLAS:2009zsq}. Using then eqs.~(\ref{eq:deltaZexp}) and~(\ref{eq:L2}), we obtain:
\begin{multline}
\label{eq:deltaZth}
 \Delta z_{\gamma} = r_{\phi z} -\frac{p_{\gamma z}}{p_{\gamma T}^2}(r_{\phi x}\,p_{\gamma x}+r_{\phi y}\,p_{\gamma y}) \\
  + \frac{p_{\gamma z}}{p_{\gamma T}}\left(\frac{R_1 R_2}{R_2-R_1}\right)\Bigg\{
 \left(1-\frac{d_0^2}{R_1^2}\right)^{1/2}
 -\left(1-\frac{d_0^2}{R_2^2}\right)^{1/2}\Bigg\}~,
\end{multline}
written in terms of $\vec r_\phi$ and $\vec p_\gamma$ components. Here we define the photon transverse momentum as $p^2_{\gamma T}=p^2_{\gamma x}+p^2_{\gamma y}$, and $d_0=(r_{\phi x}\,p_{\gamma y}-r_{\phi y}\,p_{\gamma x})/p_{\gamma T}$ is the transverse impact parameter of the photon.

 We have compared this new method with the old one used in~\cite{Delgado:2022fea,Duarte:2023tdw}, which performs a 3D reconstruction of $|\Delta z_\gamma|$ instead of the 2D used by ATLAS. The difference is given by the last term in eq.~(\ref{eq:deltaZth}). We found that large part of the effect are washed out by the detector resolution. However, it is important to clarify that we did find events where the old $|\Delta z_\gamma|$ differed significantly with respect to the new one. Fortunately, these were not common, and did not have significant impact on the final results in~\cite{Delgado:2022fea,Duarte:2023tdw}.

Once the photon pointing and timing information was obtained, it was stored within the same \texttt{HepMC} file, and passed on to \texttt{Delphes} in order to carry out the detector simulation, including smearing.

\subsection{Event reconstruction and improved muon resolution}

The event reconstruction is simulated by \texttt{Delphes 3.5.0}~\cite{deFavereau:2013fsa}, which depends on \texttt{FastJet 3.4.0}~\cite{Cacciari:2011ma}. The detector simulation includes most modules from the ATLAS card packaged within \texttt{Delphes}, with modifications as described in~\cite{Duarte:2023tdw}. In particular, we adapted photon/electron track isolation, photon/electron calorimeter isolation, and photon/electron ID efficiencies, according to~\cite{ATLAS:2022vhr}.

In this work, we have improved the muon isolation implementation compared to~\cite{Duarte:2023tdw}, in order to apply an overlap removal algorithm closer to that in~\cite{ATLAS:2022vhr}. Our procedure for muon isolation is now based on information from the \texttt{PflowLoose} working point of section~5.3 in~\cite{ATLAS:2020auj}, as found in Chapter 6.2.2 in~\cite{Mahon:2021bai}. For definiteness, we consider tracks with transverse momentum larger that 0.5~GeV, and all energy deposits from neutral particles, within a $\Delta R=0.2$ fixed cone around each muon. The scalar sum of these tracks is labelled $p_{\rm trk}$, while that of energy deposits is $E_T^{\rm neut}$. From this, a muon with transverse momentum $p_T^\mu$ is considered as isolated if $(p_{\rm trk}+0.4E_T^{\rm neut})<0.16\,p_T^\mu$. Our procedure is implemented by appropriately modifying the corresponding \texttt{Delphes} isolation subroutine, using the \texttt{tracks} and \texttt{eflowNeutralHadrons} variables to calculate $p_{\rm trk}$ and $E_T^{\rm neut}$, respectively. Since muon isolation has an efficiency above $95\%$, we have introduced this new procedure without modifying the total identification efficiency used in~\cite{Duarte:2023tdw}.

After object reconstruction it is necessary to implement the overlap removal algorithm. According to~\cite{ATLAS:2022vhr}, if an electron or jet is found within $\Delta R\leq0.4$ of a photon, they are removed from the event. Similarly, if a jet is found within $\Delta R\leq0.2$ from an electron, it is also removed from the event. However, if an electron is found within $\Delta R\leq0.4$ from any surviving jet, then the electron is removed. Finally, any muon with $\Delta R\leq0.4$ from photons or surviving jets were removed.

The last step of this algorithm, i.e.\ favouring jets over muons, provided problematic, as \texttt{Delphes} initially classifies all muons as jets, and then performs its own overlap removal on the \texttt{UniqueObjectFinder} module, which favours muons over jets. In~\cite{Duarte:2023tdw}, an independent overlap removal was performed, but faced the problem of having all muons classified as jets by \texttt{Delphes}.  To overcome this, we modified the Isolation module in \texttt{Delphes} to return both isolated and non-isolated muons, allowing us to consider only the non-isolated ones in the jet algorithm.

\subsection{Statistical analysis}
\label{sec:statistical}

 The ATLAS search estimated the dominant background to be due to pp collisions. The measured number of events was consistent with the background expectation, implying constraints on the production of signal events. We calculate our 95\% exclusion limits using the CLs method, as done in~\cite{Duarte:2023tdw}. The upper limit to the number of signal events $s^{\rm up}$ is $6.8$ ($3.8$) for the single (multi\nobreakdash) photon channels. The search in~\cite{ATLAS:2022vhr} also presents a ``combination'' of both channels, with $s^{\rm up}= 6.8$. However, this is essentially a combination of the information from the single and multi-photon channels into a single category, which reflects the fact that both channels are maximally correlated due to the procedure for assigning the events to one or the other category.

In any case, the obtained values of $s^{\rm up}$ allow us to constrain the branching ratio of Higgs decays into $\phi$ pairs, ${\rm BR}(h\to\phi\phi)$, via:
\begin{eqnarray}\label{eq:events}
\mathcal L\,\sum_{X=Z,\,W,\,t\bar t}K_X\,\sigma_{Xh}\,{\rm BR}(h \to \phi \phi)\frac{N^{\rm cuts}_X}{N^{\rm gen}_X}\leq s^{\rm up}~,
\end{eqnarray}
where $\mathcal L$ is the integrated luminosity, $\sigma_{Xh}$ are the cross-sections for each Higgs production process, $K_X$ are the associated K-factors~\cite{LHCHiggsCrossSectionWorkingGroup:2016ypw}, $N_X^{\rm gen}$ are the number of generated events for each process, and $N_X^{\rm cuts}$ are the corresponding number of events surviving the cuts. This bound on ${\rm BR}(h \to \phi \phi)$ can also be translated into a limit on $\lambda_{HS}$, using eq.~(\ref{eq:GammaHiggs}).

\bibliographystyle{JHEP}
\bibliography{bibliography}

\end{document}